\documentclass[a4paper,preprintnumbers,showpacs,twocolumn,superscriptaddress,nofootinbib,notitlepage]{revtex4-2}

\usepackage{dsfont}
\usepackage{mathtools,leftindex,tensor,mhchem}
\usepackage{booktabs}
\usepackage{siunitx}
\usepackage{enumitem}
\usepackage{times}
\usepackage{dcolumn}
\usepackage{comment}
\usepackage{hyperref}
\usepackage{amsmath,accents}
\usepackage{mathtools}
\usepackage{bbm}
\usepackage{bm}
\usepackage{amsfonts}
\usepackage{amssymb}
\usepackage{amsthm}
\usepackage{mathrsfs}
\usepackage[scr=rsfs,cal=boondox]{mathalfa}
\usepackage{wasysym}
\usepackage{graphicx}
\usepackage[]{subfigure}
\usepackage{filecontents}
\usepackage[dvipsnames]{xcolor}
\usepackage{bbold}
\usepackage[scr=rsfs,cal=boondox]{mathalfa}

\usepackage{stackengine}
\stackMath
\newcommand\tsup[2][2]{%
 \def\useanchorwidth{T}%
  \ifnum#1>1%
    \stackon[-1.3ex]{\tsup[\numexpr#1-1\relax]{#2}}{\mathchar"307E}%
  \else%
    \stackon[-1ex]{#2}{\mathchar"307E}%
  \fi%
}
\hypersetup{
    bookmarks=true,         
    pdftoolbar=true,        
    pdfmenubar=true,        
    pdffitwindow=true,     
    pdfstartview={FitH},     
    pdftitle={My title},     
    pdfauthor={author},      
    pdfsubject={Subject},    
    pdfcreator={Creator},    
    pdfproducer={Producer},  
    pdfkeywords={keyword1} 
    pdfnewwindow=true,       
    colorlinks=true ,        
    linkcolor=Blue  ,        
    citecolor=Blue,      
    filecolor=Blue,       
    urlcolor=Blue            
}

\newcommand{\oalpha}[1]{\accentset{\circ}{\alpha}}
\newcommand{\obf}[1]{\accentset{\circ}{\mathbf{f}}}
\newcommand{\boR}[1]{\accentset{\circ}{\mathbf{R}}}
\newcommand{\obF}[1]{\accentset{\circ}{\mathbf{F}}}
\newcommand{\obPi}[1]{\accentset{\circ}{\mathbf{\Pi}}}

\usepackage{scalerel}
\usepackage{tikz}
\usetikzlibrary{svg.path}

\definecolor{orcidlogocol}{HTML}{A6CE39}
\tikzset{
  orcidlogo/.pic={
    \fill[orcidlogocol] svg{M256,128c0,70.7-57.3,128-128,128C57.3,256,0,198.7,0,128C0,57.3,57.3,0,128,0C198.7,0,256,57.3,256,128z};
    \fill[white] svg{M86.3,186.2H70.9V79.1h15.4v48.4V186.2z}
                 svg{M108.9,79.1h41.6c39.6,0,57,28.3,57,53.6c0,27.5-21.5,53.6-56.8,53.6h-41.8V79.1z M124.3,172.4h24.5c34.9,0,42.9-26.5,42.9-39.7c0-21.5-13.7-39.7-43.7-39.7h-23.7V172.4z}
                 svg{M88.7,56.8c0,5.5-4.5,10.1-10.1,10.1c-5.6,0-10.1-4.6-10.1-10.1c0-5.6,4.5-10.1,10.1-10.1C84.2,46.7,88.7,51.3,88.7,56.8z};
  }
}

\newcommand\orcidicon[1]{\href{https://orcid.org/#1}{\mbox{\scalerel*{
\begin{tikzpicture}[yscale=-1,transform shape]
\pic{orcidlogo};
\end{tikzpicture}
}{|}}}}

\begin{document}

\title{\textbf{Charged Simpson-Visser AdS Black Holes: Geodesic Structure and Thermodynamic Properties }}

\author{Faizuddin Ahmed\orcidicon{0000-0003-2196-9622}
}
\email{faizuddinahmed15@gmail.com}
\affiliation{Department of Physics, The Assam Royal Global University, Guwahati 781035, Assam, India\\
}

\author{Ahmad Al-Badawi\orcidicon{0000-0002-3127-3453}
}
\email{ahmadbadawi@ahu.edu.jo}
\affiliation{Department of Physics, Al-Hussein Bin Talal University, 71111, Ma'an, Jordan\\
}
\author{Mohsen Fathi\orcidicon{0000-0002-1602-0722}}
\email{mohsen.fathi@ucentral.cl (Corresponding author)}
\affiliation{Centro de Investigaci\'{o}n en Ciencias del Espacio y F\'{i}sica Te\'{o}rica (CICEF), Universidad Central de Chile, La Serena 1710164, Chile\\
}

\begin{abstract}

In this article, we apply the Simpson-Visser (SV) regularization scheme to Anti-de Sitter (AdS) charged black holes and investigate the resulting spacetime geometry in detail, with emphasis on both geodesic structure and thermodynamic behavior. In particular, we analyze the motion of massless particle, focusing on key features such as the photon sphere, black hole shadow, photon trajectory and the dynamics of charged particles, including the characteristics of the circular and type of orbits. Furthermore, we compare the theoretical predictions of the charged SV-AdS black hole with recent observations reported by the Event Horizon telescope (EHT) for M87* and Sgr~A*. Beyond the geodesic analysis, we explore the thermodynamics of the regularized charged SV-AdS black hole by deriving essential quantities such as the Hawking temperature, Gibbs free energy, and specific heat capacity. Through a systematic examination of these thermodynamic variables, we demonstrate how the regularization parameter inherent in the SV regularization influences particle dynamics, stability conditions, and the overall thermal properties of the modified black hole solution. This comprehensive study highlights the interplay between regularization effects and the physical observables associated with charged AdS black holes.
\bigskip

{\noindent{\textit{keywords}}: Black holes; modified theories of gravity; anti-de Sitter space; geodesics; thermodynamics}\\


\end{abstract}

\date{\today}
\maketitle

\section{Introduction}

\noindent The study of black holes has gained remarkable attention in recent years, driven by groundbreaking advances in observational astronomy. The EHT collaboration has produced the first direct images of a black hole shadow, initially capturing M87* in 2019 \cite{EHTL1,EHTL4,EHTL6} and, more recently, imaging Sagittarius A* (Sgr A*), the supermassive black hole at the center of our galaxy \cite{EHTL12,EHTL25,EHTL26}. These observations provide compelling empirical confirmation of the existence of black holes as predicted by general relativity and open unprecedented opportunities to probe fundamental physics under extreme gravitational conditions. The image of M87* revealed a bright emission ring encircling a central dark shadow, consistent with theoretical models of a Kerr black hole, thereby enabling precise tests of gravity in the vicinity of the event horizon. Likewise, the imaging of Sgr A* offers crucial insights into the dynamics and environment of our galactic center. These experimental achievements underscore the importance of theoretical and phenomenological studies of black holes-including their geometry, thermodynamics, and interactions with surrounding matter fields-to deepen our understanding of their astrophysical significance and fundamental properties. Such investigations are vital not only for interpreting observational data but also for advancing our knowledge of gravitational physics, quantum effects near horizons, and the underlying nature of spacetime.  

Black holes are arguably the most ideal thermal systems in nature and have played a central role in deepening our understanding of general relativity, quantum mechanics, information theory, and statistical physics. The formulation of the four laws of black hole mechanics \cite{Bardeen1973,Wald2001} formalized their macroscopic behavior and suggested a close connection to thermodynamic properties. Building on the pioneering work of Bekenstein, Hawking demonstrated that black holes can emit radiation due to quantum effects near the event horizon \cite{Hawking1975,Hawking1976}. This radiation, known as \emph{Hawking radiation}, arises from vacuum fluctuations and can be interpreted as a tunneling process, resulting in a small but measurable flux of energy.  These insights established that black holes possess well-defined thermodynamic quantities such as entropy and temperature \cite{Bekenstein1972,Bekenstein1973}. In particular, the Hawking temperature is proportional to the surface gravity at the event horizon, and black holes obey a set of thermodynamic laws analogous to those of conventional thermodynamic systems. Together, these results reveal that black holes can be treated as genuine thermodynamic systems, providing a unique bridge between gravity, quantum mechanics, and statistical physics.   

Black holes in general relativity are singular solutions of the Einstein equations. According to Penrose singularity theorem \cite{Penrose,Penrose1}, singularities characterized by geodesic incompleteness are inevitable if appropriate energy conditions hold. Such singular behavior, however, undermines predictability, motivating the search for modifications that incorporate quantum effects at high curvatures \cite{kiefer01,kiefer02}. Even within a classical framework, several proposals exist to regularize black hole spacetimes. The first notable example was Bardeen's model \cite{Bardeen}, where the asymptotic mass is replaced by an \(r\)-dependent function to render curvature invariants finite, resulting in a black hole with a regular center but a standard event horizon. These \emph{regular black holes} have since been extensively studied \cite{reg2,reg3,reg4,reg5,reg6}. Simpson {\it et al.} introduced a minimalistic regularization scheme \cite{SV}, producing a spacetime that can describe either a regular black hole or a traversable wormhole depending on the choice of the free parameter. In this construction, all curvature invariants, including the Kretschmann and Weyl scalars, remain finite throughout, though the associated stress-energy tensor violates the null energy condition, as is typical for regular black holes.  \\     

To date, the SV-regularization scheme has not been applied to charged AdS black holes in the literature. An uncharged SV-regular AdS black hole was recently explored in Ref. \cite{Kumar2025}, highlighting the influence of the regularization parameter on horizon properties. Extending this framework to charged AdS black holes is particularly compelling, as the presence of charge coupled with the regularization parameter is expected to induce significant modifications in the thermal behavior and phase structure of the black hole. In the present work, we implement the SV-regularization procedure for charged AdS black hole spacetimes and systematically investigate their physical properties. We study the dynamics of massless and charged particles, with a focus on the photon sphere, black hole shadow, effective radial forces experienced by photons and photon trajectories. Furthermore, we analyze the thermodynamic properties of the black hole, emphasizing its phase structure. In the standard Reissner-Nordstr\"{o}m-AdS (RN-AdS) scenario , the Hawking-Page transition describes a shift from thermal AdS to the black hole phase as the Hawking temperature varies. It is therefore of considerable interest to examine how the SV-regularization modifies this phase structure. In this study, we assume that the standard laws of black hole thermodynamics continue to hold for SV-regular black holes, allowing us to explore the interplay between regularization, charge, and AdS effects in a unified framework.  \\   

The paper is organized as follows: In  section  \ref{sec:2}  we introduce the charged SV-AdS black hole. Section \ref{sec:null} presents a detailed examination of geodesic motions, analyzing  null trajectories, effective potentials, black hole shadow and constraints from the EHT, and angular velocity and effective radial force of photons. Section \ref{sec:charged12} represents dynamics of charged test particles including circular 
and types of orbits. Section \ref{sec:therom} explores the thermodynamic properties of the black hole, deriving key quantities such as Hawking temperature, entropy, and specific heat to characterize thermal behavior and stability. Sections \ref{sec:therom2} and \ref{sec:therom3} discuss modified first law of thermodynamics and equation of state respectively. Finally, in Section \ref{conclu}, we summarize our findings.   \\

\section{SV-AdS Regular Charged Black Holes} \label{sec:2}

Simpson and Visser devised a simple regularization procedure \cite{SV} with one parameter such that it lifted the spacetime singularity and rendered any black hole solution with a regular geometry everywhere. According to this procedure, in the Schwarzschild coordinates (\(t, r, \theta, \phi \)), the coordinate \(r\) is replaced by \(\sqrt{r^2+a^2}\). Here, \(a\) is some real parameter which we shall refer to as SV-regularization parameter in the present article. This leads to extending the domain of \(r\) from \(r\in (0,+\infty)\) to \(r\in (-\infty,+\infty)\) for some nonzero value of \(a\). A Schwarzschild black hole metric with this regularization is given by 
\begin{equation}
    ds^2=-f(r)dt^2+\frac{dr^2}{f(r)}+h^2(r)\,d\Omega^2,\label{aa1}
\end{equation}
where
\begin{align}
    f(r)=1-\frac{2M}{\sqrt{r^2+a^2}},\qquad h^2(r)=r^2+a^2\,.\label{aa2}
    \end{align}
Here, \(M\) is the ADM mass and \(d\Omega^2\) is a metric on unit 2-sphere. The black hole horizon radius is given by \(f(r_h)=0\), that is, \(r_h=\pm\sqrt{(2M)^2-a^2}\). It is clear that the metric represents a regular black hole for \(a< 2M\) \cite{SV}. For \(a\geq2M\), the geometry becomes a wormhole.

The RN-AdS black hole, which generalizes the Schwarzschild-AdS solution by incorporating a nonvanishing electric charge $Q$, is characterized by the electromagnetic four-potential $A_{\mu} = -Q \delta^{t}_{\mu}/r$. The corresponding spacetime geometry is determined by the lapse function \cite{Wu2000}
\begin{align}
    f(r)=1-\frac{2M}{r}+\frac{Q^2}{r^2}+\frac{r^2}{\ell^2_p},\qquad h^2(r)=r^2.\label{aa3}
    \end{align}
Following the regularization procedure employed in the SV model, we introduce a modified lapse function suitable for describing a charged AdS regular black hole. This construction preserves the essential physical features of the original solution while removing the central singularity through an appropriate deformation of the metric function. By incorporating the effects of electric charge and the negative cosmological constant within the AdS background, the resulting lapse function provides a smooth, well-behaved geometry across the entire spacetime. This approach ensures that the metric remains finite at the core and transitions seamlessly to the expected RN-AdS behavior at large radial distances, thereby offering a consistent framework for exploring regular black hole solutions in asymptotically AdS spacetimes. The lapse function is given by
\begin{align}
    f(r)&=1-\frac{2M}{\sqrt{r^2+a^2}}+\frac{Q^2}{r^2+a^2}+\frac{r^2+a^2}{\ell^2_p},\nonumber\\
    h^2(r)&=r^2+a^2,\label{aa4}
\end{align}
where \(\ell_p\) is the AdS radius. It is related to the cosmological constant as $\Lambda=-{3}/{\ell_p^2}$ spacetime dimensions. The electromagnetic potential in the current case will be $A_{\mu}=-Q \delta^{t}_{\mu}/{h(r)}$.


In this article, we are mostly interested in the geodesic structure and thermodynamic properties of such spacetime geometries. These aspects are discussed in the forthcoming sections.

\section{The Null Geodesic Structure}\label{sec:null}

The optical appearance of a black hole is determined by the structure of null geodesics in its gravitational field.  
Two key features emerge from this analysis: the photon sphere and the resulting shadow.  
The mechanism by which photons are either captured or escape to infinity in Schwarzschild geometry was first clarified by Synge~\cite{Synge1966}, who showed that unstable circular null orbits define a critical impact parameter.  
This critical threshold, associated with the photon sphere, specifies the apparent boundary of the shadow and governs the behavior of strongly lensed trajectories, a point extensively examined in both early and modern studies~\cite{Luminet1979,VirbhadraEllis2000}.  

The significance of these optical structures has increased considerably following the EHT observations of M87* and Sgr A*, which provided the first horizon-scale image of supermassive black holes~\cite{EHT2019I,event_horizon_telescope_collaboration_first_2022}.  
As a result, photon spheres and shadows now serve as powerful tools for probing general relativity and for constraining alternative gravitational frameworks.

\begin{figure*}[t]
    \centering
    \includegraphics[width=8cm]{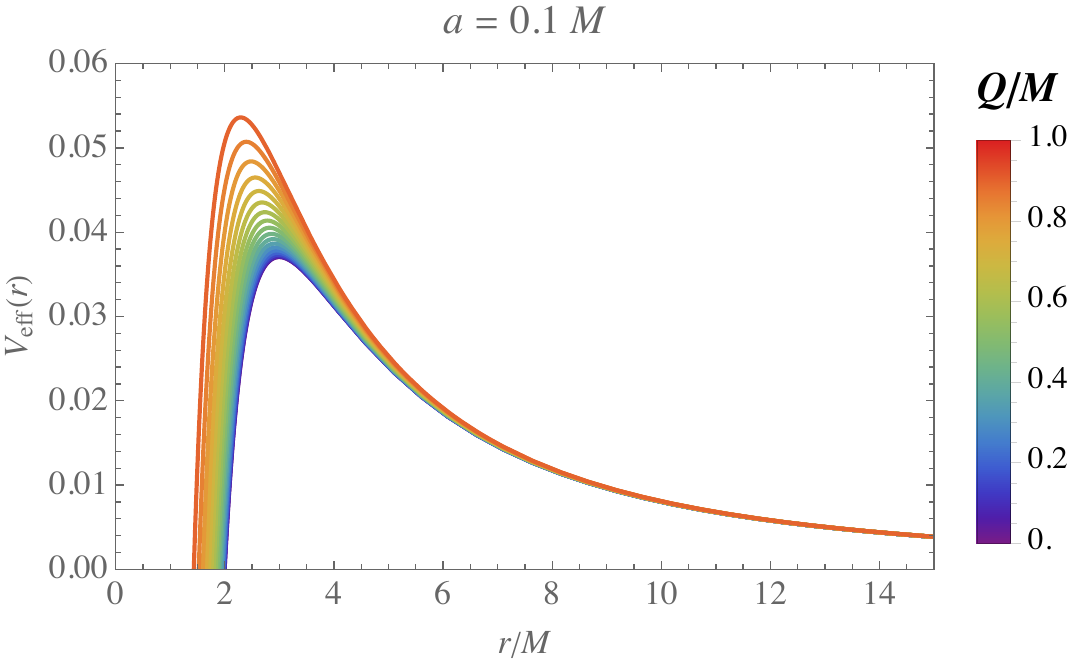} (a)\quad
    \includegraphics[width=8cm]{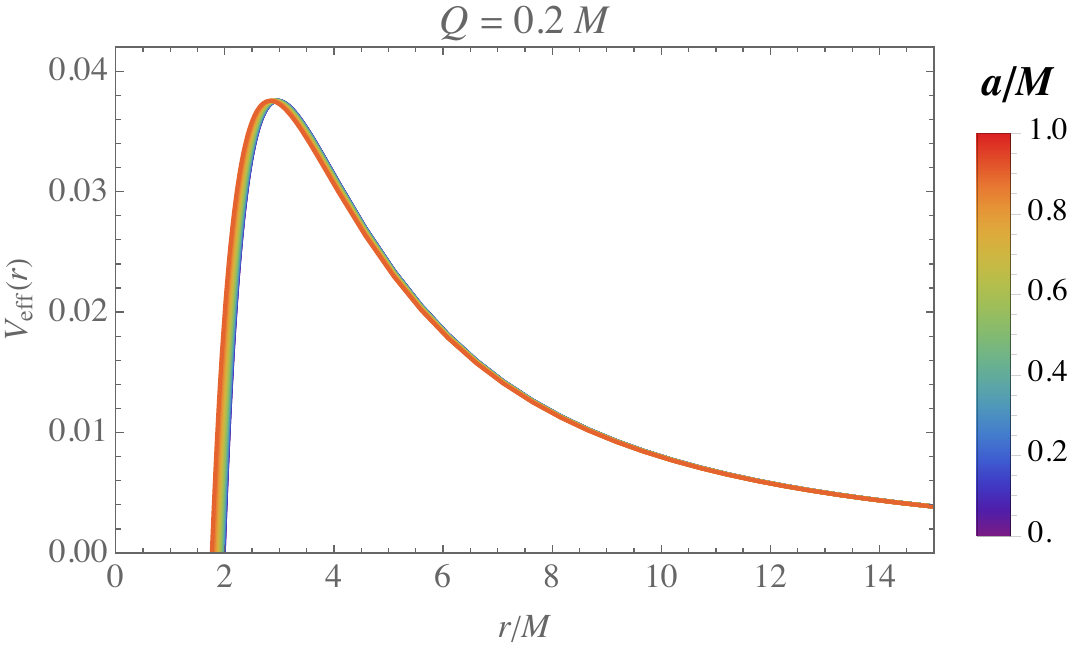} (b)
    \caption{The profiles of $V_\mathrm{eff}(r)$, plotted for $\ell_p=500 M$ and $L = 1 M$, versus (a) changes in $Q$ while $a = 0.1 M$, and (b) changes in $a$ while $Q=0.2 M$.}
    \label{fig:Veff_gen}
\end{figure*}

To analyze the propagation of light in any black hole background, one begins with the null geodesics that describe the motion of massless particles through the curved spacetime.  
These trajectories determine the bending of light in the strong-field region and thereby the shape and extent of the shadow.  
In the case of a charged SV-AdS spacetime, the dynamics of photon orbits follow from the Lagrangian associated with the underlying metric (see, for example, Refs. \cite{ALBADAWI2025185,AHMED2025101925,AHMED2025116951,AHMED2025101988}).  
This Lagrangian encodes the geometric structure and relevant physical contributions of the model, and it forms the starting point for deriving the equations that govern photon motion and the resulting shadow configuration.  
It is written as
\begin{equation}
\mathbb{L} = \frac{1}{2} g_{\mu\nu} \dot{x}^\mu\dot{x}^\nu,
    \label{bb1}
\end{equation}
in which an overdot denotes differentiation with respect to the affine parameter, i.e., $d/d\lambda$, where $\lambda$ is the affine parameter along the null worldlines. Without loss of generality, the photon trajectories can be restricted to the equatorial plane by setting $\theta=\pi/2$.  
Under this assumption, the Lagrangian takes the form
\begin{equation}
    \mathbb{L}=\frac{1}{2}\left[-f(r)\,\dot{t}^{2}
    +\frac{\dot{r}^{2}}{f(r)}
    +h^2(r)\dot{\phi}^{2}\right].
    \label{bb2}
\end{equation}
The conjugate momenta then follow from
\begin{equation}
    \Pi_{\mu}=\frac{\partial \mathbb{L}}{\partial \dot{x}^{\mu}},
    \label{eq:PI}
\end{equation}
which yield two conserved quantities associated with the Killing symmetries of the spacetime:
\begin{eqnarray}
    \Pi_{t} &=& -f(r)\,\dot{t}\;\doteq\; \mathrm{E}, \label{bb3}\\
    \Pi_{\phi} &=& h^2(r)\,\dot{\phi}\;\doteq\; \mathrm{L}. \label{bb4}
\end{eqnarray}
We interpret $\mathrm{E}$ and $\mathrm{L}$ as the photon's energy and angular momentum, respectively.  
These constants naturally define the impact parameter \cite{bozza_gravitational_2010}
\begin{equation}
    b=\frac{\mathrm{L}}{\mathrm{E}} = \frac{h^2(r)}{f(r)}\frac{d\phi}{d t},
    \label{eq:b_def}
\end{equation}
which plays a central role in determining the boundary of the black hole shadow. For photon trajectories, the condition $\mathbb{L}=0$ governs the dynamics, which leads to the equations
\begin{eqnarray}
    && \dot{r}^{2} + V_{\rm eff}(r) = \mathrm{E}^{2},
    \label{bb5}\\
    && \dot\phi=\frac{\mathrm{L}}{h^2(r)},\label{bb5a}
\end{eqnarray}
where the effective potential governing the motion of incoming photons is given by
\begin{equation}
    V_{\rm eff}(r)
    = \mathrm{L}^{2}\frac{f(r)}{h^{2}(r)}. 
    \label{bb6}
\end{equation}
In Fig.~\ref{fig:Veff_gen}, we present the radial profile of the effective potential for different values of the spacetime parameters $Q$ and $a$.
The turning points $r_{t}$ of the photon trajectories are determined by the condition $\dot{r}=0$, which is equivalent to requiring
\begin{equation}
    V_{\mathrm{eff}}(r_{t}) = \mathrm{E}^{2}.
\end{equation}
The effective potential $V_{\mathrm{eff}}(r)$ admits a local maximum at the radius $r_{p}$ of the unstable photon orbit.  
At this point, the standard extremal conditions
\begin{equation}
    V_{\mathrm{eff}}(r_{p}) = \mathrm{E}_{p}^{2},
    \qquad
    V_{\mathrm{eff}}'(r_{p}) = 0,
\end{equation}
must be satisfied, where primes denote derivatives with respect to $r$. This maximum can be hence obtained as
\begin{equation}
r_p = \sqrt{R_p^2-a^2},
\label{eq:rp}
\end{equation}
where
\begin{equation}
R_p = \frac{3M+\sqrt{9M^2-8Q^2}}{2},
    \label{eq:R}
\end{equation}
is the radius of the photon sphere for the RN black hole. The condition $9M^{2} - 8Q^{2} \ge 0$ (equivalently $Q^{2} \le \tfrac{9}{8} M^{2}$) ensures that the quantity $R_{p}$ remains real, i.e.\ $R_{p} \in \mathbb{R}$.  
In addition, the requirement $R_{p}^{2} > a^{2}$ must hold so that the associated radius $r_{p}$ is real and positive.

The critical impact parameter, defined as $b_{p} \equiv \mathrm{L}_{p}/\mathrm{E}_{p}$, follows from Eq. \eqref{eq:b_def}, reading
\begin{equation}
    b_{p}^{2} \equiv b^2(r_p) = \frac{h^{2}(r_{p})}{f(r_{p})}
              = \frac{R_{p}^{2}}{f(r_{p})},
    \label{eq:bp0}
\end{equation}
which leads to
\begin{equation}
    b_{p}
    = \frac{R_{p}}
           {\displaystyle
            \sqrt{
            1 - \frac{2M}{R_{p}}
            + \frac{Q^{2}}{R_{p}^{2}}
            + \frac{R_{p}^{2}}{\ell_{p}^{2}}
            }}.
    \label{eq:bp1}
\end{equation}
Several limiting cases are worth highlighting.  
When $a = 0$, one has $r_{p} = R_{p}$; in particular, setting $Q = 0$ recovers the well-known Schwarzschild value $r_{p} = 3M$.  
In the opposite regime, where $a \gg r$, the geometry of the photon surface becomes dominated by $a$, and the photon sphere eventually disappears as $a$ grows sufficiently large.  
For $a \le 0$ with $Q = 0$, the radius reduces to $r_{p} = \sqrt{(3M)^{2} - a^{2}}$, showing that the photon sphere shrinks as $a$ increases in magnitude.  
More generally, for small deviations with $a \ll M$, the photon sphere radius admits the expansion
\begin{equation}
    r_{p} \approx R_{p}
                 - \frac{a^{2}}{6M}
                 + \mathcal{O}(a^{4}),
    \label{eq:rp_a}
\end{equation}
which quantifies the leading correction induced by the parameter $a$.

By introducing the change of variable $r \doteq 1/u$, the angular equation of motion can be derived from Eqs.~\eqref{bb5} and \eqref{bb5a}, yielding
\begin{equation}
\left(\frac{du}{d\phi}\right)^2 = P(u),
\label{eq:du/dphi0}
\end{equation}
where
\begin{eqnarray}
P(u) &=&
\left(\frac{1}{b^2} - \frac{1}{\ell_p^2}\right)
+ u^2\left(\frac{2a^2}{b^2} - 1 - \frac{2a^2}{\ell_p^2}\right)\nonumber\\
&& + 2 M u^3 \sqrt{1 + a^2 u^2}\nonumber\\
&& + u^4\left(\frac{a^4}{b^2} - a^2 - Q^2 - \frac{a^4}{\ell_p^2}\right).
\label{eq:P_0}
\end{eqnarray}
In the limit $a \rightarrow 0$, the above expression reduces to the fourth-order polynomial
\begin{equation}
\tilde{P}_4(u) = \frac{1}{b^2} - u^2 + 2 M u^3 - Q^2 u^4 - \frac{1}{\ell_p^2}\,,
\label{eq:P4_RNAdS}
\end{equation}
which coincides with the corresponding result for the RN-(A)dS spacetime.  
It is straightforward to verify that the photon sphere is determined by the simultaneous conditions $P(u) = 0$ and $P'(u) = 0$, which uniquely fix the critical impact parameter $b = b_p$.

The function in Eq.~\eqref{eq:P_0} determines the turning points $u_t$ of photon trajectories approaching the black hole as the real solutions of $P(u_t) = 0$. Since $u_p \equiv 1/r_p$ corresponds to the maximum of the effective potential, photons with $u_t = u_d < u_p$ (i.e.\ $r_d > r_p$ and $b > b_p$) are scattered by the black hole, giving rise to relativistic lensed images. In contrast, for $u_t = u_f > u_p$ (i.e.\ $r_f < r_p$ and $b < b_p$), photons inevitably fall into the event horizon. Consequently, the black hole shadow is formed by the set of photon trajectories that asymptotically approach the photon sphere at $r = r_p$.

\begin{figure*}
    \centering
    \includegraphics[width=7cm]{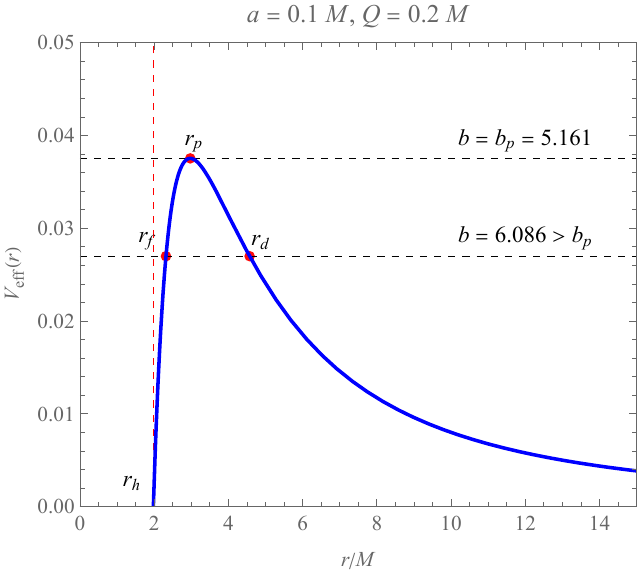} \qquad
    \includegraphics[width=7cm]{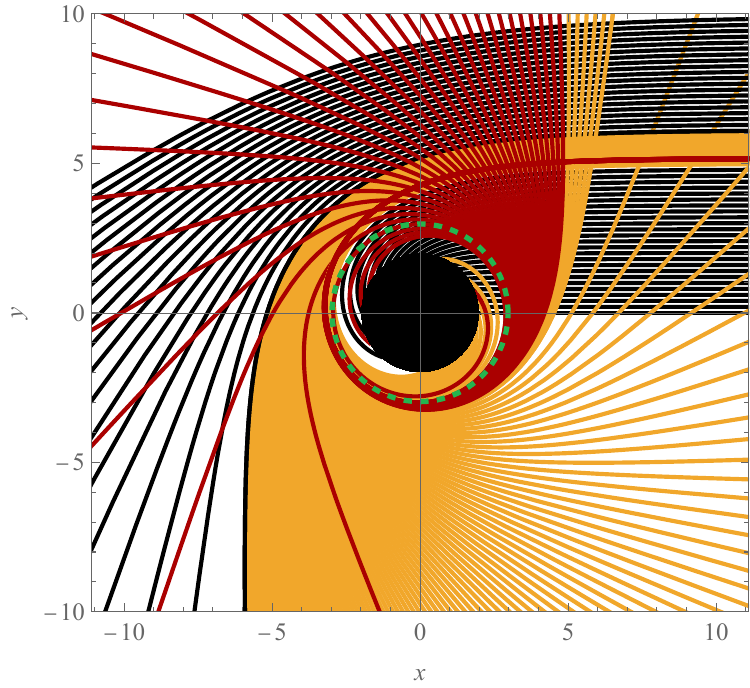}
    \caption{A typical radial profile of the effective potential is shown together with the corresponding shape of equatorial null geodesic orbits plotted in the for specific values of the spacetime parameters $a$ and $Q$, assuming $L = 1 M$ and $\ell_p = 500 M$. The characteristic radii for this case are: $r_h = 1.977 M$, $r_d = 4.572 M$, $r_f = 2.318 M$, and $r_p = 2.971 M$. In the orbit diagram, the black disk represents the event-horizon radius $r = r_h$, while the dashed green circle denotes the photon-sphere radius $r = r_p$. Furthermore, the black, yellow, and red trajectories correspond to photons executing zero, one, and two half-orbits around the black hole, respectively, before being deflected away from or captured by the black hole.}
    \label{fig:Veff}
\end{figure*}

In Fig.~\ref{fig:Veff}, a representative profile of the effective potential $V_{\mathrm{eff}}(r)$ has been displayed, together with its corresponding turning points for different values of the impact parameter.
In this figure, we also illustrate the manner in which photons orbit the black hole during both their approach to and recession from it by plotting the corresponding null geodesics in the $(x,y)$ plane.

\subsection{Weak deflection angle in the perturbative regime}
\label{subsec:weak_deflection}

The weak deflection limit corresponds to light rays propagating far from the photon sphere, where spacetime curvature effects are small and the impact parameter $b$ is much larger than the gravitational radius of the compact object.  
In this regime, the deflection angle can be computed perturbatively by expanding the null geodesic equation in inverse powers of the impact parameter.

\begin{figure*}[t]
    \centering
    \includegraphics[width=8.2cm]{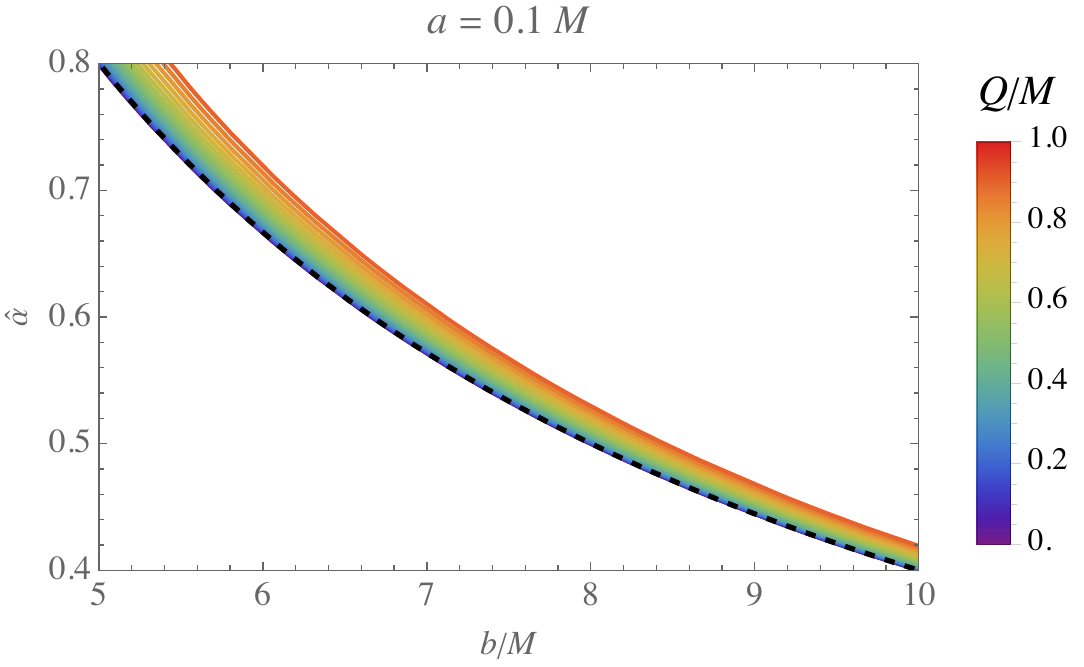} (a)\qquad
    \includegraphics[width=8.2cm]{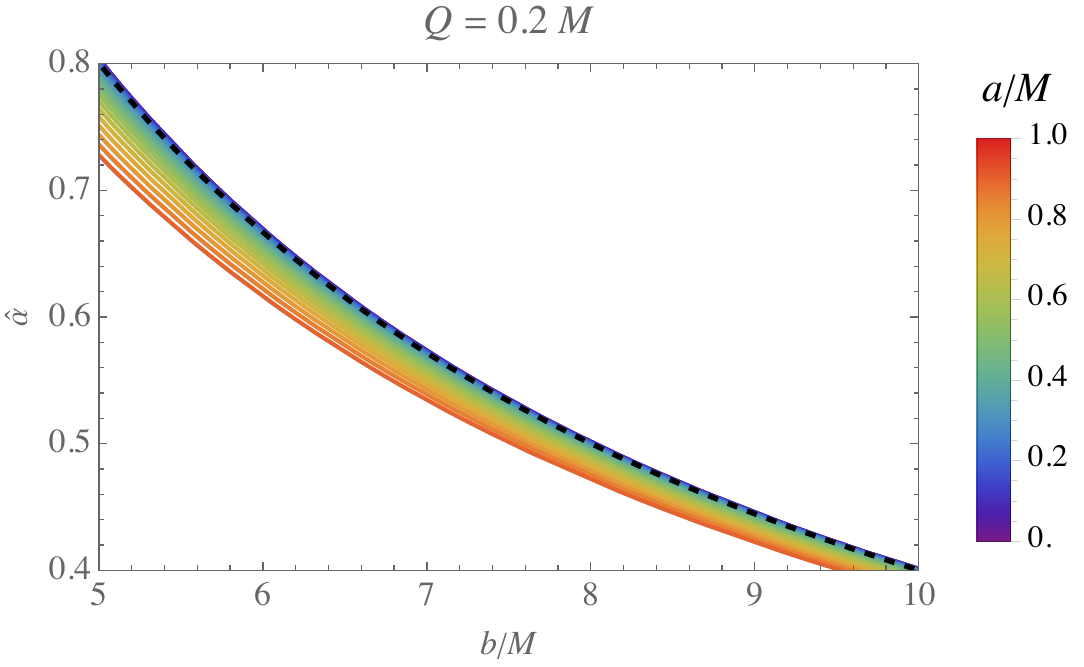} (b)
    \caption{The $b$-profiles of the weak deflection angle, plotted for $\ell_p = 500 M$, for changes of (a) $Q$ while $a=0.1 M$, and (b) $a=0.1 M$ while $Q=0.2 M$. The black dashed curve in the diagrams, corresponds to the weak deflection angle in the Schwazschild-AdS spacetime.}
    \label{fig:weak_def}
\end{figure*}

In this context, the polynomial $P(u)$ in Eq. \eqref{eq:P_0} admits the expansion
\begin{equation}
P(u) = \frac{1}{b^2} - u^2 + 2M u^3 - (Q^2 - a^2)u^4 + \mathcal{O}(u^5).
\label{eq:P_u}
\end{equation}
The weak deflection regime is characterized by the conditions
\begin{equation}
Mu \ll 1, \qquad |Q|u \ll 1, \qquad |a|u \ll 1,
\end{equation}
which allows us to solve the orbit equation perturbatively.  
Accordingly, we expand the solution as
\begin{equation}
u(\phi) = u_0(\phi) + u_1(\phi) + \mathcal{O}(M^2, Q^2, a^2),
\end{equation}
where $u_0$ is the zeroth–order (flat spacetime) solution. At leading order, Eq.~\eqref{eq:P_u} reduces to
\begin{equation}
\left( \frac{du_0}{d\phi} \right)^2 = \frac{1}{b^2} - u_0^2 ,
\end{equation}
whose solution is
\begin{equation}
u_0(\phi) = \frac{1}{b}\sin\phi .
\end{equation}
Now by substituting $u = u_0 + u_1$ into the orbit equation and keeping only linear terms, one finds
\begin{equation}
\frac{d^2 u_1}{d\phi^2} + u_1 = 3M u_0^2 - 2(Q^2 - a^2)u_0^3 .
\label{eq:u1_eq}
\end{equation}
Using $u_0 = b^{-1}\sin\phi$, the source terms become
\begin{subequations}
\begin{align}
& 3M u_0^2 = \frac{3M}{2b^2}\left(1 - \cos 2\phi\right),\\
& 2(Q^2 - a^2)u_0^3 = \frac{2(Q^2 - a^2)}{b^3}
\left(\frac{3\sin\phi - \sin 3\phi}{4}\right).
\end{align}
\end{subequations}
Hence, one can solve Eq.~\eqref{eq:u1_eq}, whose the physically relevant particular solution reads
\begin{eqnarray}
u_1(\phi) &=&
\frac{3M}{2b^2}
+ \frac{M}{2b^2}\cos 2\phi
- \frac{3(Q^2 - a^2)}{4b^3}\phi\cos\phi\nonumber\\
&&- \frac{(Q^2 - a^2)}{16b^3}\sin 3\phi .
\end{eqnarray}
Accordingly, the total deflection angle $\hat{\alpha}$ is obtained by evaluating the asymptotic deviation of the trajectory from a straight line. At large distances, the condition $u(\phi) \to 0$ yields
\begin{equation}
\phi = \pi + \hat{\alpha},
\end{equation}
and retaining the secular contribution proportional to $\phi\cos\phi$ in $u_1$, one finds
\begin{equation}
\hat{\alpha}
= \frac{4M}{b}
+ \frac{3\pi}{4}\frac{(Q^2 - a^2)}{b^2}
+ \mathcal{O}\!\left(\frac{M^2}{b^2}\right).
\label{eq:weak_deflection}
\end{equation}
The first term reproduces the standard Schwarzschild deflection angle, while the second term encodes corrections arising from the charge parameter $Q$ and the regularization scale $a$.  
Notably, the parameter $a$ enters the weak deflection angle at the same perturbative order as $Q^2$, indicating that its effect is subdominant but potentially observable in high–precision lensing measurements.
The validity of this expansion requires $b \gg r_p$, where $r_p$ denotes the photon sphere radius, ensuring that the light ray remains in the weak–field region throughout its trajectory. In Fig. \ref{fig:weak_def}, the behavior of the weak deflection angle \eqref{eq:weak_deflection} has been plotted for changes in the black hole parameters. 
%

\subsection{Black hole shadow and constraints from the EHT}\label{subsec:shadow}

The photon sphere of a black hole is characterized by unstable circular null trajectories that arise due to the strong curvature of spacetime.  
Photons following such orbits are delicately balanced between capture by the event horizon and escape to infinity, giving rise to a bright photon ring.  For a distant observer, this ring marks the outer boundary of the dark region commonly identified as the black hole shadow \cite{Synge:1966,Cunningham:1972,Bardeen:1973a,Luminet:1979nyg}. 
Early investigations described this phenomenon using a variety of terms, including the \textit{escape cone} \cite{Synge:1966}, the \textit{cone of gravitational radiation capture} \cite{zeldovich_relativistic_1966}, as well as notions such as the \textit{optical appearance of black holes} and the \textit{black hole image} \cite{Bardeen:1973b,Chandrasekhar:1998,Luminet:1979nyg,luminet_seeing_2018}.  The modern terminology \textit{black hole shadow}, introduced by Falcke \textit{et al.}~\cite{Falcke_2000}, has since become standard and refers specifically to the dark silhouette enclosed by the photon ring \cite{johannsen2010,johnson_universal_2020}. 
In this subsection, we analyze the properties of the black hole shadow and examine how its characteristics depend on the parameters of the underlying gravitational model, namely $Q$ and $a$.  

To determine the radius of the shadow cast by the charged SV-AdS black hole, we consider an observer located at a finite radial position $r_O$.  
The angular radius of the shadow, denoted by $\alpha_{\mathrm{sh}}$, is defined by those null geodesics that asymptotically approach the photon sphere before reaching the observer.  
For a general spherically symmetric spacetime described by Eq.~\eqref{aa1}, this angle satisfies the relation
\begin{equation}
\sin\alpha_{\mathrm{sh}} =
\frac{b_p}{\displaystyle \sqrt{\frac{h^2(r_O)}{f(r_O)}}}
=
\frac{b_p\,\sqrt{f(r_O)}}{\sqrt{r_O^2+a^2}},
\label{eq:alpha_sh}
\end{equation}
where $b_p$ denotes the critical impact parameter associated with the photon sphere and is given explicitly in Eq.~\eqref{eq:bp1}.

The physical radius of the shadow on the observer’s sky is then defined as
\begin{equation}
R_{\mathrm{sh}} = r_O \sin\alpha_{\mathrm{sh}},
\label{eq:Rsh_def}
\end{equation}
which provides a direct connection between the local angular size and the geometrical properties of the spacetime.  
In the limit of a distant observer, namely $r_O \gg a$, this expression simplifies to
\begin{equation}
R_{\mathrm{sh}}(r_O) \simeq b_p\,\sqrt{f(r_O)},
\label{eq:Rsh_1}
\end{equation}
recovering the familiar asymptotic relation for the shadow radius.

\begin{figure*}[t]
    \centering
    \includegraphics[width=8cm]{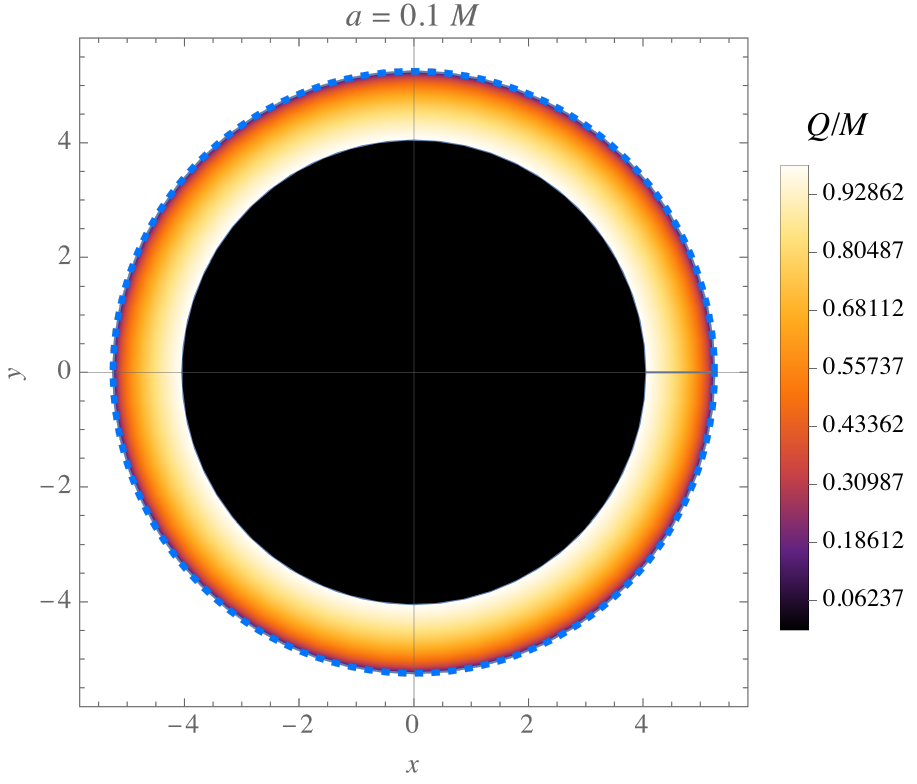}\,(a)\qquad
    \includegraphics[width=8cm]{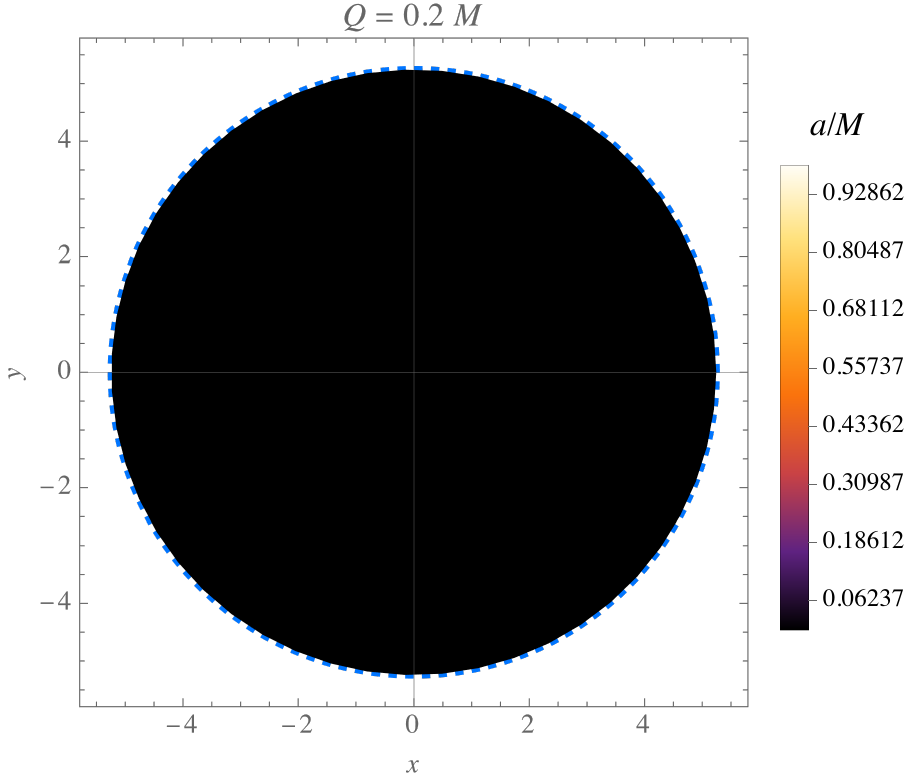}\,(b)
    \caption{Variation of the theoretical shadow radius $R_{\mathrm{sh}}$ for an observer located at $r_O = 100 M$ with $\ell_p = 500 M$:  
    (a) as a function of $Q$ for fixed $a = 0.1 M$, and  
    (b) as a function of $a$ for fixed $Q = 0.2 M$.  
    The dashed blue circle represents the shadow radius of the Schwarzschild–AdS black hole, corresponding to $a = Q = 0$, for which $R_{\mathrm{sh}}^{\mathrm{SAdS}} = 5.248 M$.  
    The central dark disk indicates the limiting configuration reached as the parameters attain larger values.}
    \label{fig:Rsh}
\end{figure*}

In Fig.~\ref{fig:Rsh}, we illustrate the dependence of the shadow radius $R_{\mathrm{sh}}$ on variations of the spacetime parameters $Q$ and $a$, while keeping the observer fixed at $r_O$.
As inferred from Fig.~\ref{fig:Rsh}, variations in the parameter $a$ induce only mild changes in the shadow radius, whereas modifications in $Q$ lead to more pronounced deviations.  
Moreover, among the configurations considered, the SA-dS black hole exhibits the largest shadow radius, highlighting the shrinking effect introduced by the additional spacetime parameters.

It is also instructive to confront the theoretical predictions of the charged SV-AdS black hole with the most recent observations reported by the EHT for the supermassive black holes M87* and Sgr~A*.  
To this end, we first recall that the theoretical diameter of the shadow is defined as
$d_{\mathrm{sh}}(r_O) = 2\,R_{\mathrm{sh}}(r_O)$. On the observational side, the shadow diameter can be inferred from the measured angular diameter $\theta_*$ through the relation~\cite{bambi_testing_2019}
\begin{equation}
d_{\mathrm{sh}} = \frac{r_O\,\theta_*}{M_*},
\label{eq:dsh_obs}
\end{equation}
where $M_*$ denotes the mass of the black hole.  

For M87*, the EHT observations correspond to a mass $M_* = (6.5 \pm 0.90)\times10^9\,M_\odot$, a distance $r_O = 16.8\,\mathrm{Mpc}$, and an angular diameter $\theta_* = 42 \pm 3\,\mathrm{\mu as}$.  
Similarly, for Sgr~A*, the reported values are $M_* = (4.3 \pm 0.013)\times10^6\,M_\odot$, $r_O = 8.127\,\mathrm{kpc}$, and $\theta_* = 48.7 \pm 7\,\mathrm{\mu as}$.  
Using Eq.~\eqref{eq:dsh_obs}, these measurements yield the shadow diameters
$d_{\mathrm{sh}}^{\mathrm{M87^*}} = 11 \pm 1.5$ and
$d_{\mathrm{sh}}^{\mathrm{SgrA^*}} = 9.5 \pm 1.4$.

\begin{figure*}[t]
    \centering
    \includegraphics[width=7cm]{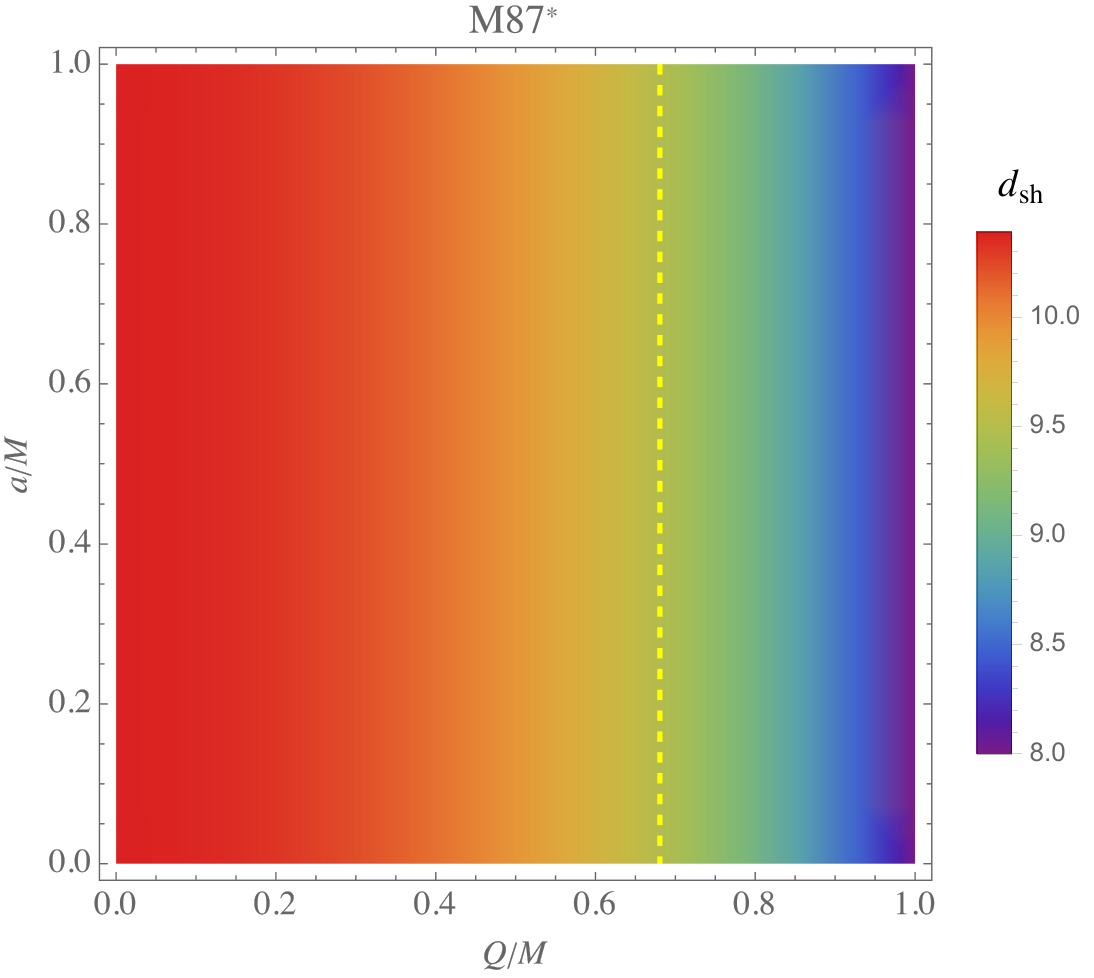}\qquad
    \includegraphics[width=7cm]{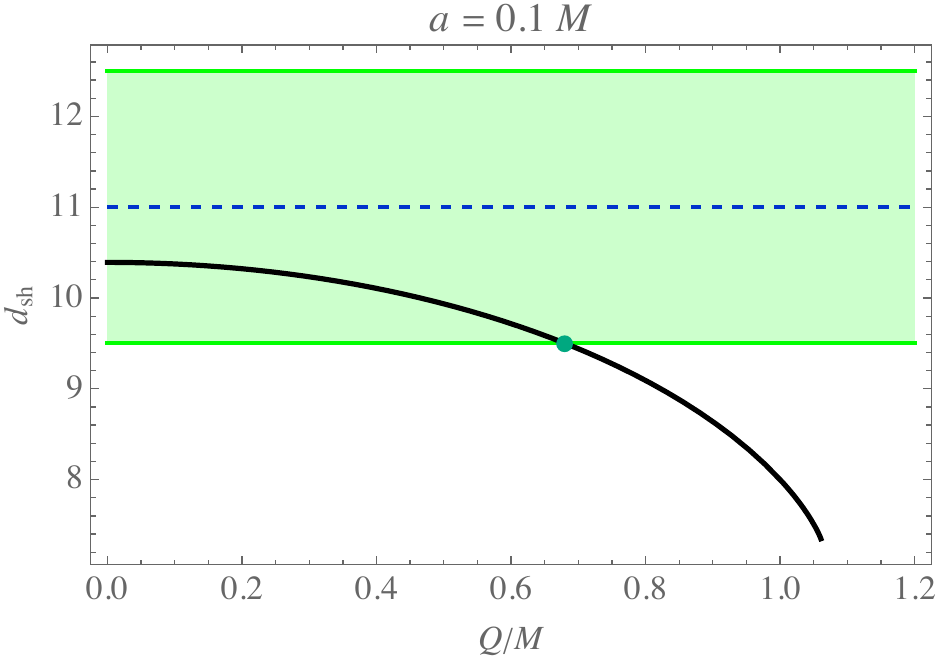}\\[2mm]
    \includegraphics[width=7cm]{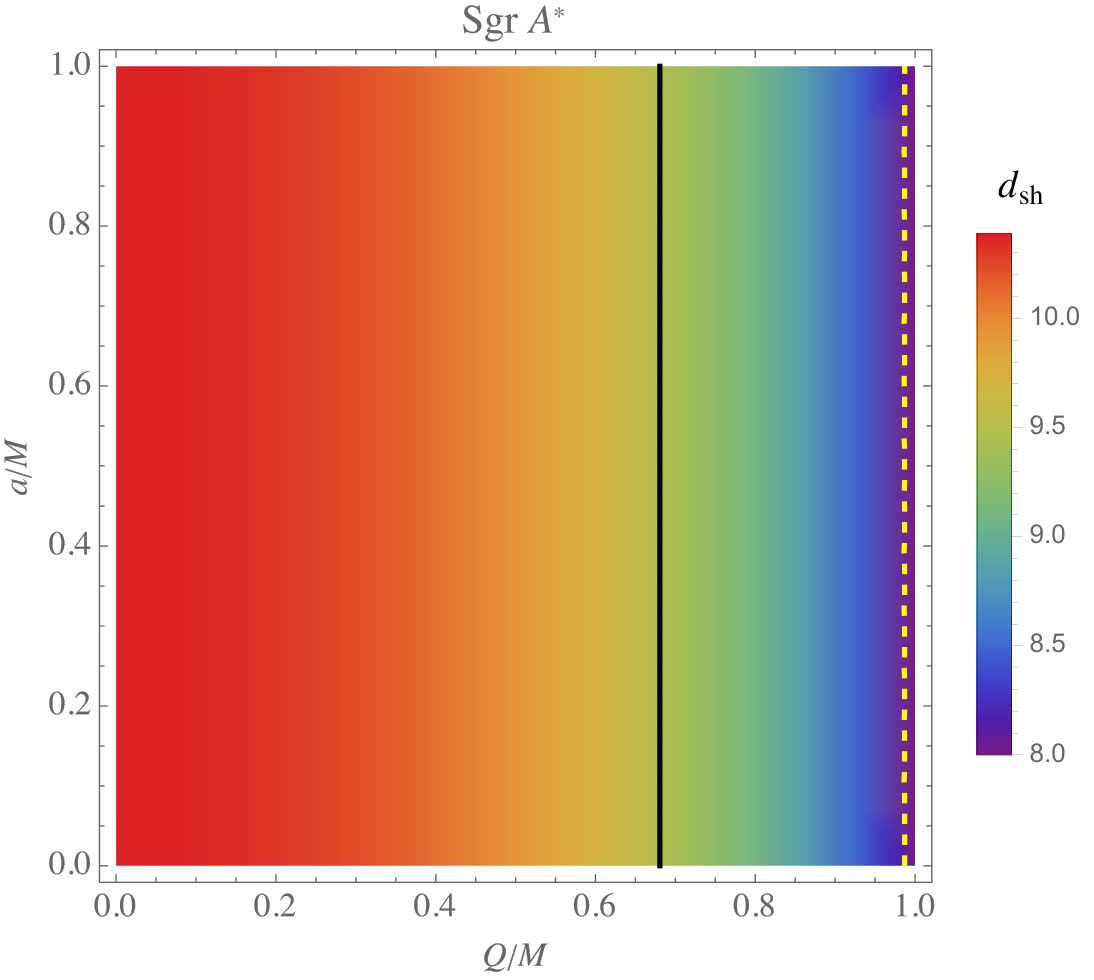}\qquad
    \includegraphics[width=7cm]{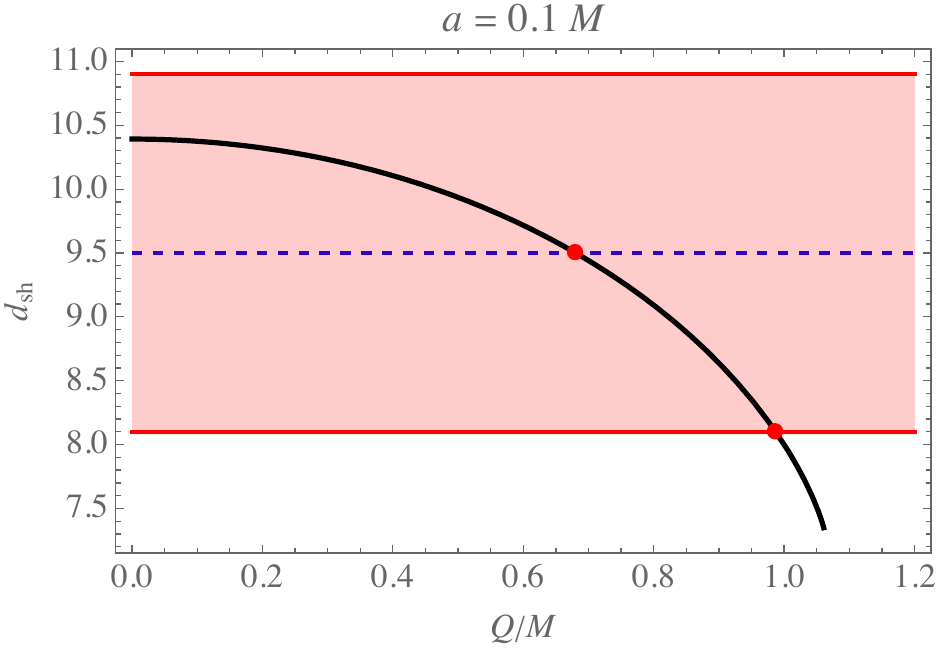}
    \caption{
    Comparison between the theoretical shadow diameter of the charged SV-AdS black hole and the EHT observations of (top row) M87* and (bottom row) Sgr~A*, assuming $\ell_p = 1.65 \times 10^{26}\,\mathrm{m}$, corresponding to $\Lambda = -1.1 \times 10^{-52}\,\mathrm{m}^{-2}$.  
    In the left panels, both parameters $a$ and $Q$ are varied simultaneously, whereas in the right panels the parameter $a$ is fixed and the dependence on $Q$ is displayed explicitly.  
    For M87*, the yellow dashed line in the left panel marks the value $d_{\mathrm{sh}} = 9.5$, which corresponds to $Q \simeq \pm 0.681\,M$ and coincides with the point where the black curve in the right panel intersects the $-1\sigma$ boundary of the observational confidence interval.  
    For Sgr~A*, the yellow dashed line in the left panel indicates $d_{\mathrm{sh}} = 8.1$, while the solid black line corresponds to $d_{\mathrm{sh}} = 9.5$.  
    These values translate to $Q \simeq \pm 0.987\,M$ and $Q \simeq \pm 0.681\,M$, respectively, marking the intersections of the theoretical curve with the $-1\sigma$ bound and the central observational value in the right panel.
    }
    \label{fig:EHT_const}
\end{figure*}

In Fig.~\ref{fig:EHT_const}, we compare these observationally inferred diameters with the theoretical predictions of the charged SV-AdS black hole, assuming that either M87* or Sgr~A* is described by this spacetime.  
The dependence of the shadow diameter on the model parameters $a$ and $Q$ is explored in two complementary ways.  
In the left panels, both parameters are allowed to vary simultaneously, while in the right panels the parameter $a$ is held fixed and only $Q$ is varied.
From Fig.~\ref{fig:EHT_const}, we find that the charge-like parameter $Q$ is tightly constrained, with $|Q| \lesssim 0.7 M$. In contrast, the parameter $a$ produces only minor corrections to the shadow size and remains weakly constrained within the observational uncertainties. Breaking this degeneracy would require additional observables beyond the shadow diameter, such as photon ring substructure or lensing signatures.

\subsection{Angular velocity and effective radial force of photons}\label{subsec:angular_speed}

We now examine two complementary dynamical quantities associated with photon motion in the given spacetime: the angular velocity of photons on circular null orbits and the effective radial force governing their stability.

\subsubsection{Angular velocity}  

For photons moving along circular null trajectories with constant radius $r=\mathrm{const.}$ in the equatorial plane, the angular velocity in the azimuthal direction is defined as
\begin{equation}
    \omega^{\rm null}_{\phi}
    =
    \left.\frac{\dot{\phi}}{\dot{t}}\right|_{r=\mathrm{const.}}
    =
    \left.\frac{\mathrm{L}}{\mathrm{E}}\frac{f(r)}{h^{2}(r)}\right|_{r=\mathrm{const.}}.
    \label{eq:omega_def}
\end{equation}
Evaluating this expression at the radius $r=r_p$ of a circular null orbit and using the photon sphere condition, the angular velocity reduces to
\begin{equation}
    \omega^{\rm null}_{\phi}
    =
    \frac{\sqrt{f(r_p)}}{h(r_p)}
    .
    \label{eq:omega_rp}
\end{equation}
This quantity characterizes the orbital frequency of photons trapped near the photon surface and is directly related to observational features such as photon rings and strong lensing signatures.

\begin{figure*}[t]
    \centering
    \includegraphics[width=8.2cm]{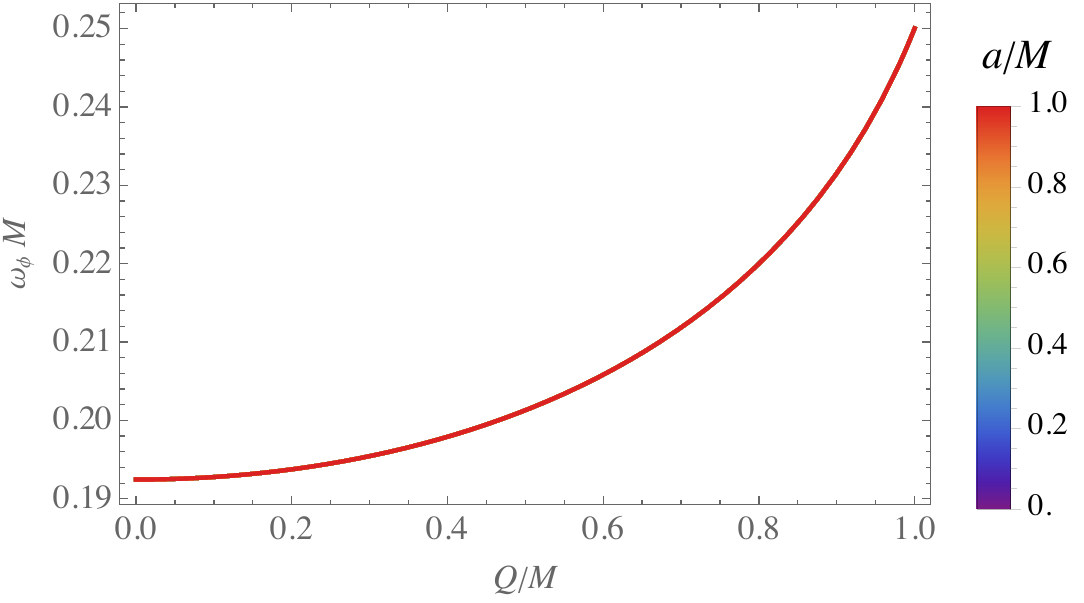} (a)\qquad
    \includegraphics[width=8.2cm]{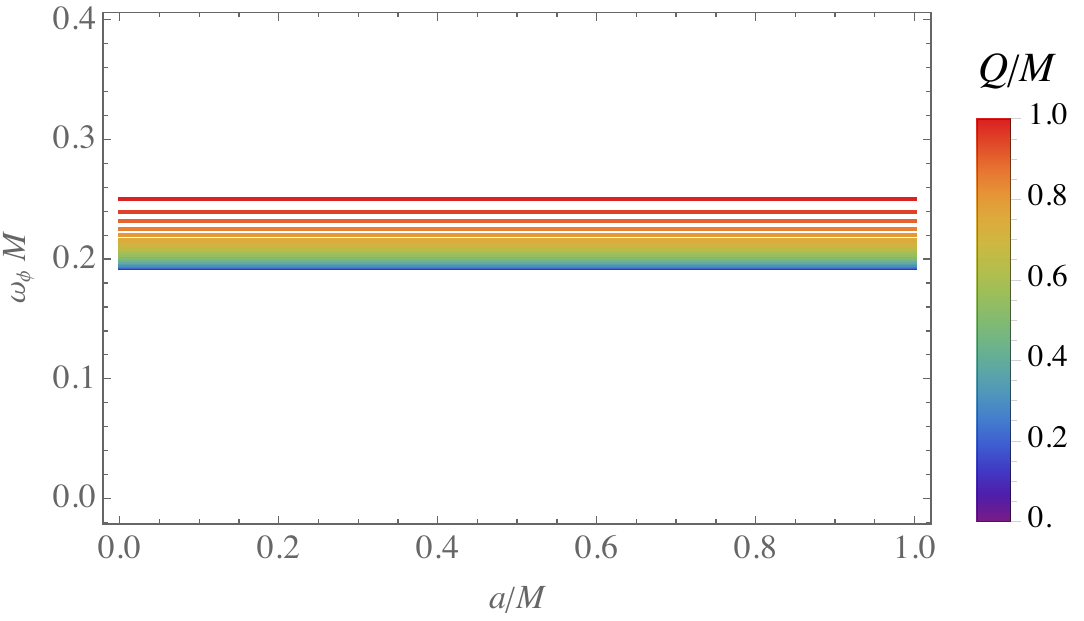} (b)
    \caption{Response of the angular velocity $\omega_\phi$ to variations of the black hole parameters, shown for $\ell_p = 500 M$.}
    \label{fig:omega_phi}
\end{figure*}

In Fig.~\ref{fig:omega_phi}, we illustrate the dependence of the angular velocity $\omega_\phi$ on variations of the black hole parameters.
As inferred from the plots, the angular velocity exhibits a pronounced sensitivity to changes in the charge parameter $Q$.  
In contrast, variations of the parameter $a$ produce a negligible effect on $\omega_\phi$ over the range considered, indicating that the rotational dynamics of circular photon orbits are predominantly governed by the charge contribution rather than by the deformation parameter $a$.

\subsubsection{Effective radial force}  

To further understand the stability of photon trajectories, we introduce the effective radial force experienced by photons in the gravitational field.  
This force determines whether nearby photon trajectories are driven outward toward infinity or inward toward the black hole.

It is defined as minus one half of the radial derivative of the effective potential for null geodesics,
\begin{equation}
    \mathrm{F}_{\mathrm{eff}}
    =
    -\frac{1}{2}\frac{d V_{\mathrm{eff}}(r)}{dr}.
    \label{eq:Feff_def}
\end{equation}
Substituting the effective potential given in Eq.~(\ref{bb6}) and simplifying, one finds
\begin{equation}
    \mathrm{F}_{\mathrm{eff}}
    =
    \frac{r\,\mathrm{L}^{2}}{(r^{2}+a^{2})^{2}}
    \left[
    1
    -
    \frac{3M}{\sqrt{r^{2}+a^{2}}}
    +
    \frac{2Q^{2}}{r^{2}+a^{2}}
    \right].
    \label{eq:Feff_final}
\end{equation}
The vanishing of $\mathrm{F}_{\mathrm{eff}}$ identifies the location of circular photon orbits, while the sign of its radial derivative determines their stability. In particular, a negative slope at the photon radius signals the instability of the photon sphere, a key feature underlying the formation of black hole shadows.

\begin{figure*}[t]
    \centering
    \includegraphics[width=8.2cm]{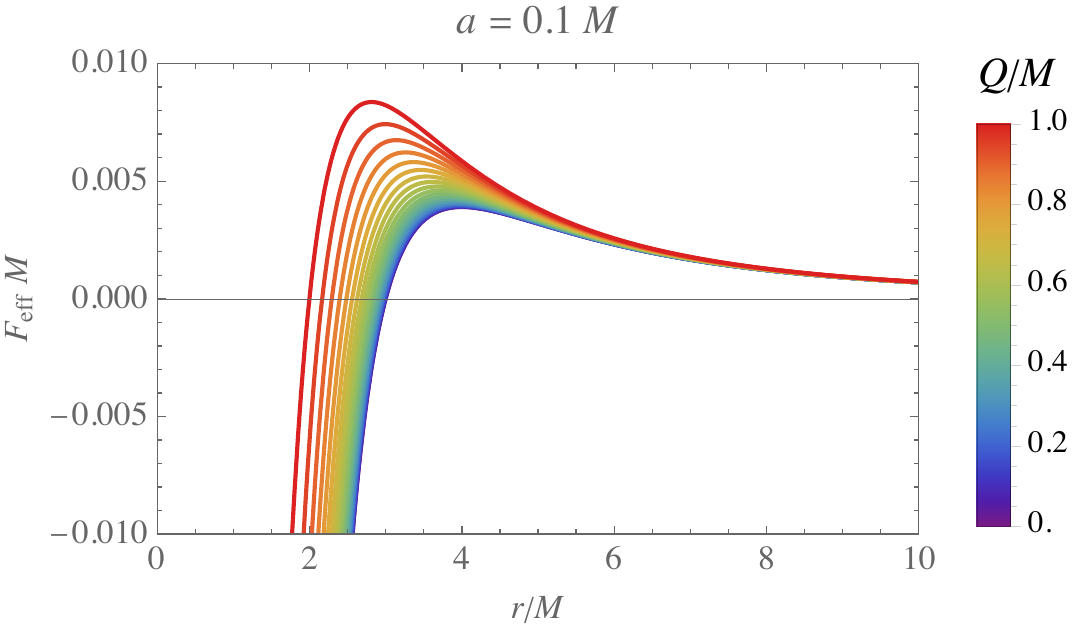} (a)\qquad
    \includegraphics[width=8.2cm]{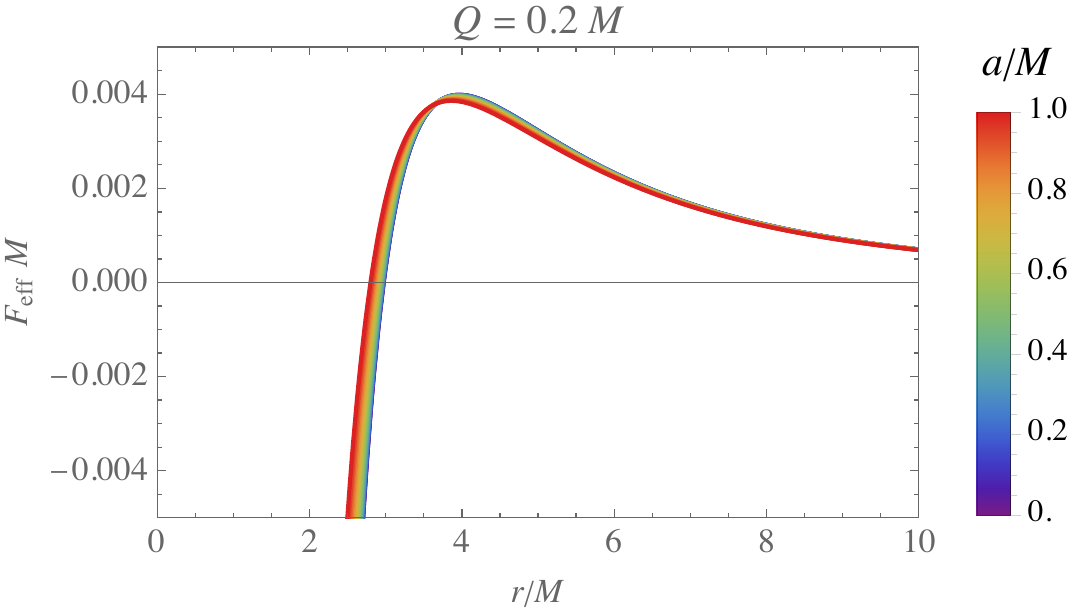} (b)
    \caption{Radial behavior of the effective force $\mathrm{F}_{\mathrm{eff}}$ for $\ell_p = 500 M$, illustrating its dependence on (a) the charge parameter $Q$ with fixed $a = 0.1 M$, and (b) the parameter $a$ with fixed $Q = 0.2 M$.}
    \label{fig:Feff}
\end{figure*}

In Fig.~\ref{fig:Feff}, we present the radial profiles of the effective force $\mathrm{F}_{\mathrm{eff}}$ for different values of the black hole parameters.
As seen from the figure, the radial profiles of $\mathrm{F}_{\mathrm{eff}}$ exhibit a single maximum in each case.  
This maximum marks the transition point where the effective radial force changes its behavior, separating regions in which photons are either driven away from the black hole or pulled inward toward the photon sphere.  
Such a feature is characteristic of unstable photon orbits and is closely related to the existence of the photon sphere.

Furthermore, the diagrams indicate that variations in the parameter $Q$ lead to noticeably stronger changes in the magnitude and location of the maximum of $\mathrm{F}_{\mathrm{eff}}$, whereas the dependence on the parameter $a$ remains comparatively weak.

\section{Dynamics of charged Test Particles}\label{sec:charged12}


As the black hole possesses an electric charge, the associated electromagnetic field is naturally described by a four--potential of the form
\begin{equation}
A_\mu\,dx^\mu = A_t(r)\,dt,
\label{eq:Amu}
\end{equation}
where the temporal component is given by
\begin{equation}
A_t(r) = -\frac{Q}{h(r)}.
\label{eq:At}
\end{equation}
In the limit $a \to 0$, this expression correctly reduces to the electromagnetic potential of the RN spacetime.

For a test particle of mass $m$ and electric charge $e$, the dynamics are governed by the action
\begin{equation}
S = \int \left( \frac{1}{2} g_{\mu\nu}\dot{x}^\mu \dot{x}^\nu + q\,A_\mu \dot{x}^\mu \right) d\lambda,
\label{eq:S_q}
\end{equation}
where $q = e/m$ denotes the specific charge of the particle and $\lambda$ is an affine parameter along the worldline.

Exploiting the stationarity and axial symmetry of the spacetime, one identifies two conserved quantities associated with the Killing vectors $\partial_t$ and $\partial_\phi$, namely the specific energy $\mathcal{E}$ and the specific angular momentum $\mathcal{L}$. These are given by
\begin{eqnarray}
\mathcal{E} &=& f(r)\dot{t} + q A_t(r),
\label{eq:EE}\\
\mathcal{L} &=& h^2(r)\dot{\phi}.
\label{eq:LL}
\end{eqnarray}
Using these constants of motion, the radial equation governing the particle trajectory can be written as
\begin{equation}
\dot{r}^2 = \left( \mathcal{E} - q A_t \right)^2 - f(r)\left[ 1 + \frac{\mathcal{L}^2}{h^2(r)} \right].
\label{eq:rrdot}
\end{equation}
It is convenient to recast this expression into a factorized form,
\begin{equation}
\dot{r}^2 = \left( \mathcal{E} - U_{\mathrm{eff}}^- \right)
            \left( \mathcal{E} - U_{\mathrm{eff}}^+ \right),
\label{eq:rrdot_1}
\end{equation}
where the effective potentials $U_{\mathrm{eff}}^\pm(r)$ are defined as
\begin{equation}
U_{\mathrm{eff}}^\pm(r) =
q A_t(r) \pm \sqrt{ f(r)\left[ 1 + \frac{\mathcal{L}^2}{h^2(r)} \right] }.
\label{eq:U+-}
\end{equation}
The effective potential governing the motion of charged test particles therefore consists of two distinct contributions: a gravitational term encoded in the metric functions and an electromagnetic (Coulomb) interaction arising from the black hole charge.  
Its qualitative behavior depends explicitly on the sign of the specific charge $q$, reflecting whether the electromagnetic interaction is attractive or repulsive.  
Throughout this work, we focus on the physically relevant branch $U_{\mathrm{eff}}(r) \equiv U_{\mathrm{eff}}^+(r)$ and refer to it simply as the effective potential.

\begin{figure*}[t]
    \centering
    \includegraphics[width=8cm]{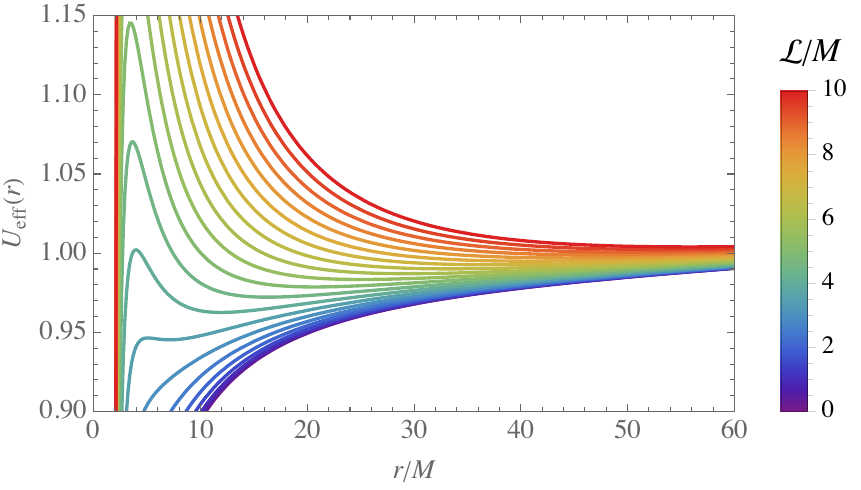} (a)\qquad
    \includegraphics[width=8cm]{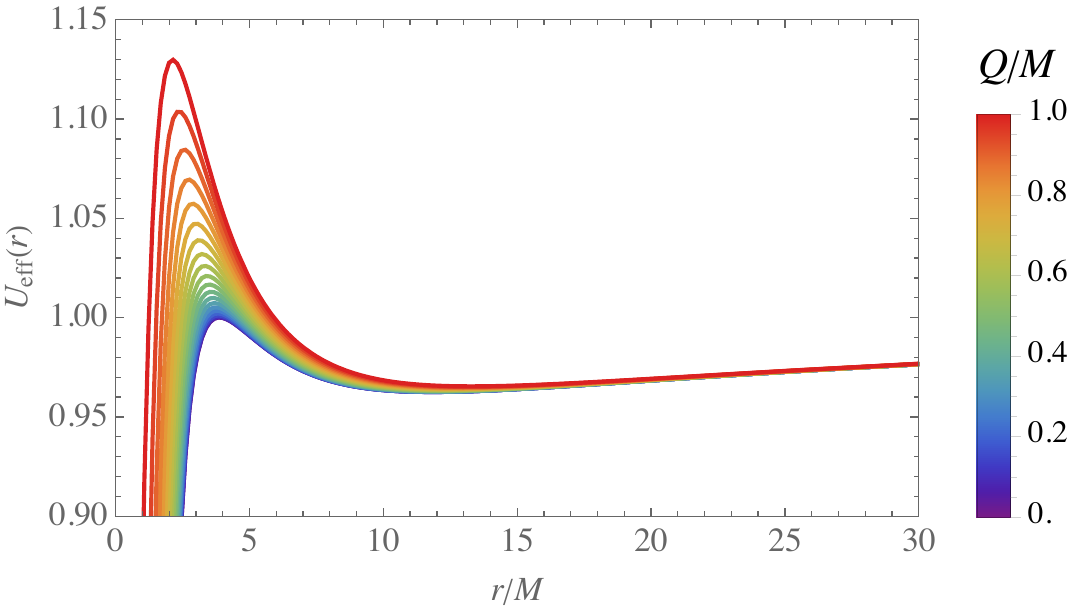} (b)
    \caption{Radial behavior of the effective potential $U_{\mathrm{eff}}(r)$ for $a=0.1M$, $\ell_p=500M$, and $q=0.01$.  
    Panel (a) shows the effect of varying $\mathcal{L}$ for fixed $Q=0.2M$, while panel (b) illustrates the impact of changing $Q$ for fixed $\mathcal{L}=4M$.}
    \label{fig:Ueff}
\end{figure*}

In Fig.~\ref{fig:Ueff}, representative radial profiles of $U_{\mathrm{eff}}(r)$ are displayed for different choices of the spacetime and particle parameters.
As inferred from the figure, increasing the angular momentum $\mathcal{L}$ at fixed $Q$ raises the effective potential barrier, thereby enlarging the classically forbidden region.  
A similar enhancement of the potential occurs when the black hole charge $Q$ is increased while keeping $\mathcal{L}$ fixed, indicating that both centrifugal and electromagnetic effects play a significant role in shaping the orbital structure of charged particles.

\subsection{Circular orbits}\label{subsec:circular}

The conditions for circular motion of charged test particles are determined by simultaneously solving  
$U_{\mathrm{eff}}(r) = \mathcal{E}$ and $U'_{\mathrm{eff}}(r) = 0$.  
The first condition ensures that the particle energy matches the effective potential at the orbital radius, while the second enforces the vanishing of the radial force.  
Together, these equations specify the radii of circular orbits and fix the corresponding conserved quantities.

Imposing these conditions leads to the following expressions for the specific angular momentum and energy of the charged particle:
\begin{widetext}
\begin{eqnarray}
\mathcal{L}_\pm^2(r) &=& 
\frac{1}{\left|h^2 f' - 2 f h'\right|^2}
\Biggl\{
f h^2 f' h' 
- h^3\Bigl[f'^2 - 2 q f A_t'^2\Bigr]
\pm q A_t' f h^2
\sqrt{
q^2 h^2 A_t'^2
- h'\Bigl(h f' - f h'\Bigr)
}
\Biggr\},
\label{eq:Lpm}
\\
\Bigl(\mathcal{E}_\pm(r) - q A_t\Bigr)^2 &=&
\frac{2 f}{h^2 \left|h f' - 2 f h'\right|^2}
\Biggl\{
2 f^2 h^2 h'^2
+ f h^3\Bigl[q^2 h A_t'^2 - f' h'\Bigr]
\nonumber\\
&&\qquad
\pm q f h^3 A_t'
\sqrt{
q^2 h^2 A_t'^2
- 2 h'\Bigl(h f' + 2 f h'\Bigr)
}
\Biggr\}.
\label{eq:Epm}
\end{eqnarray}
\end{widetext}
Here, the subscripts $\pm$ correspond to the two possible branches arising from the electromagnetic interaction, reflecting the dependence of the orbital properties on the sign of the particle’s specific charge $q$.

\begin{figure*}[t]
    \centering
    \includegraphics[width=8cm]{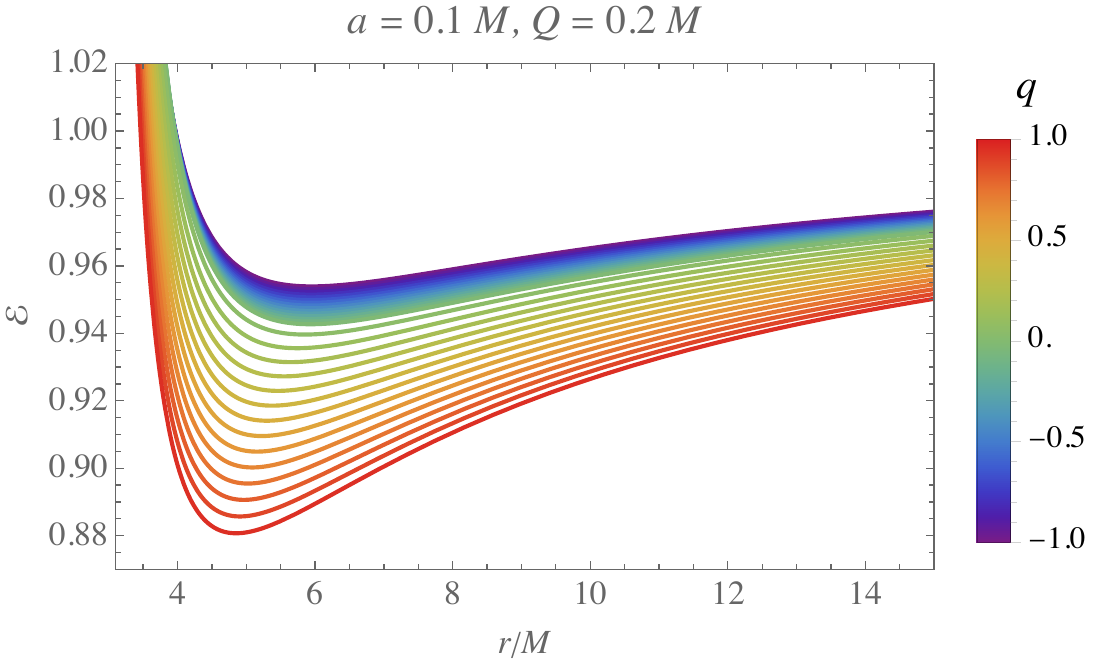} (a)\qquad
    \includegraphics[width=8cm]{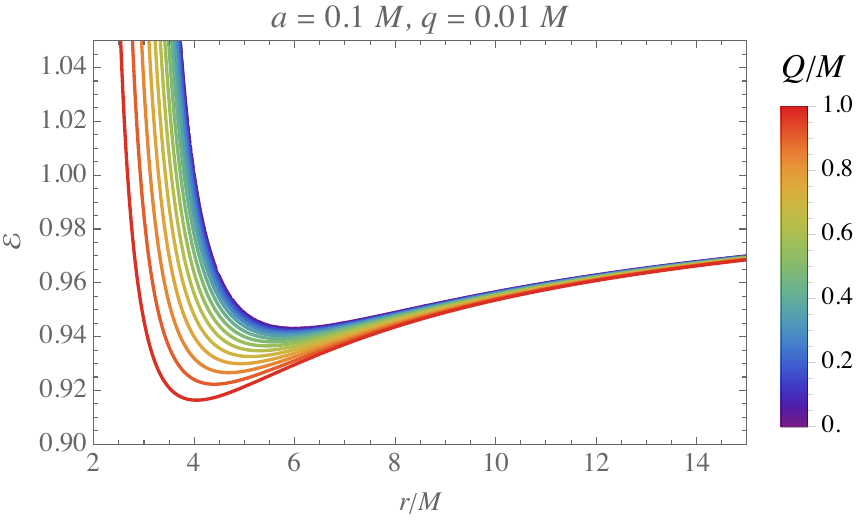} (b)
    \includegraphics[width=8cm]{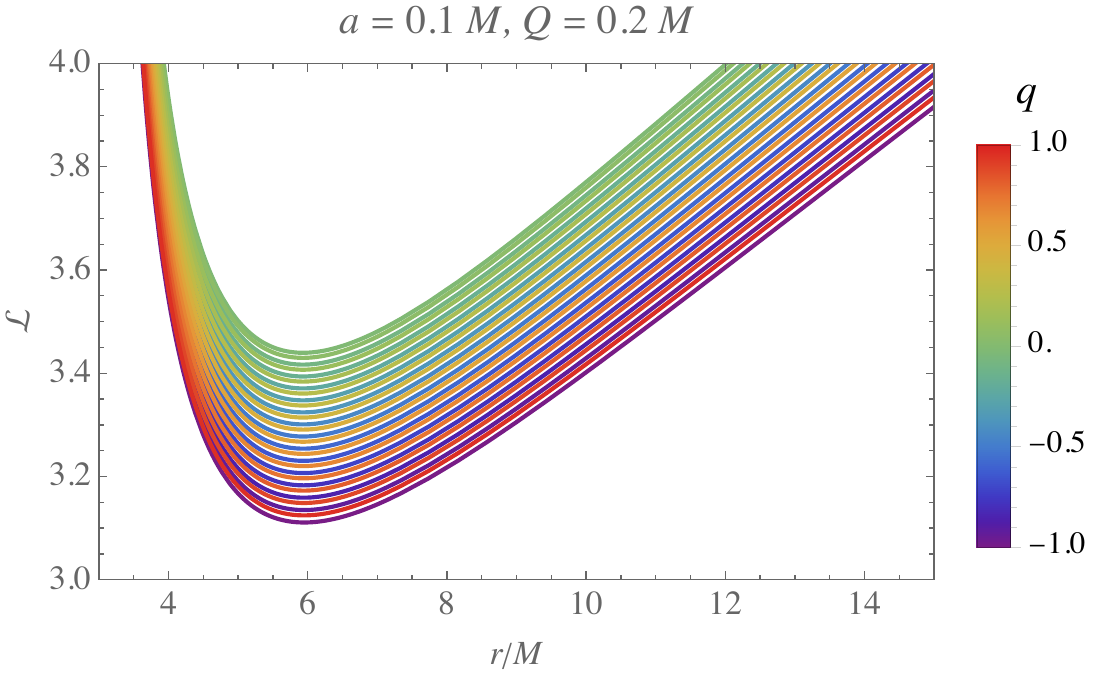} (c)\qquad
    \includegraphics[width=8cm]{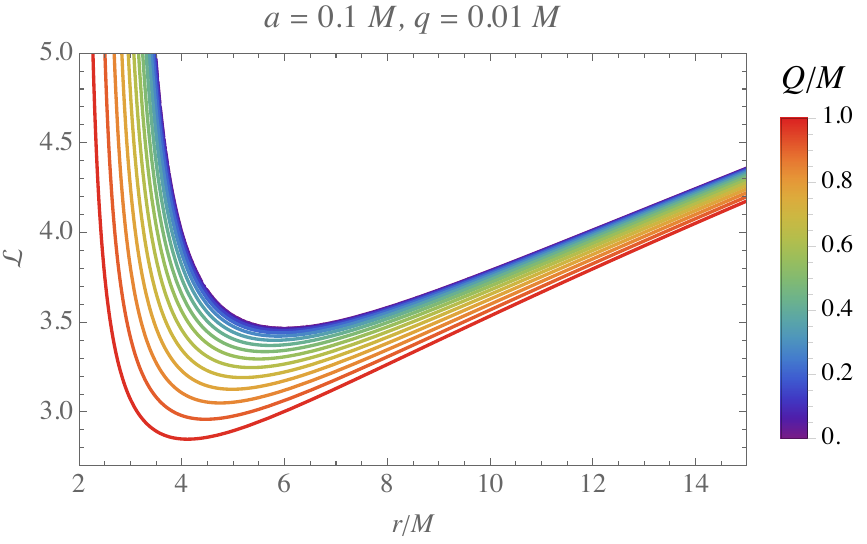} (d)
    \includegraphics[width=8cm]{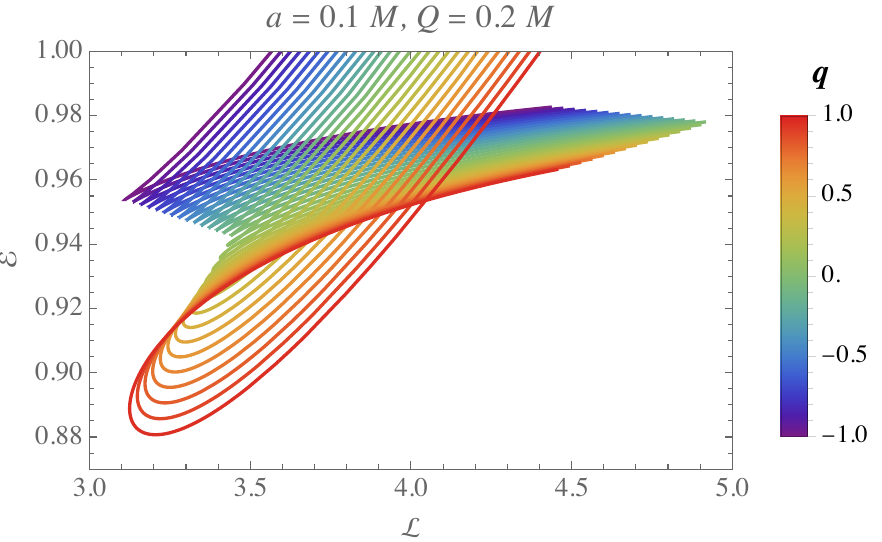} (e)\qquad
    \includegraphics[width=8cm]{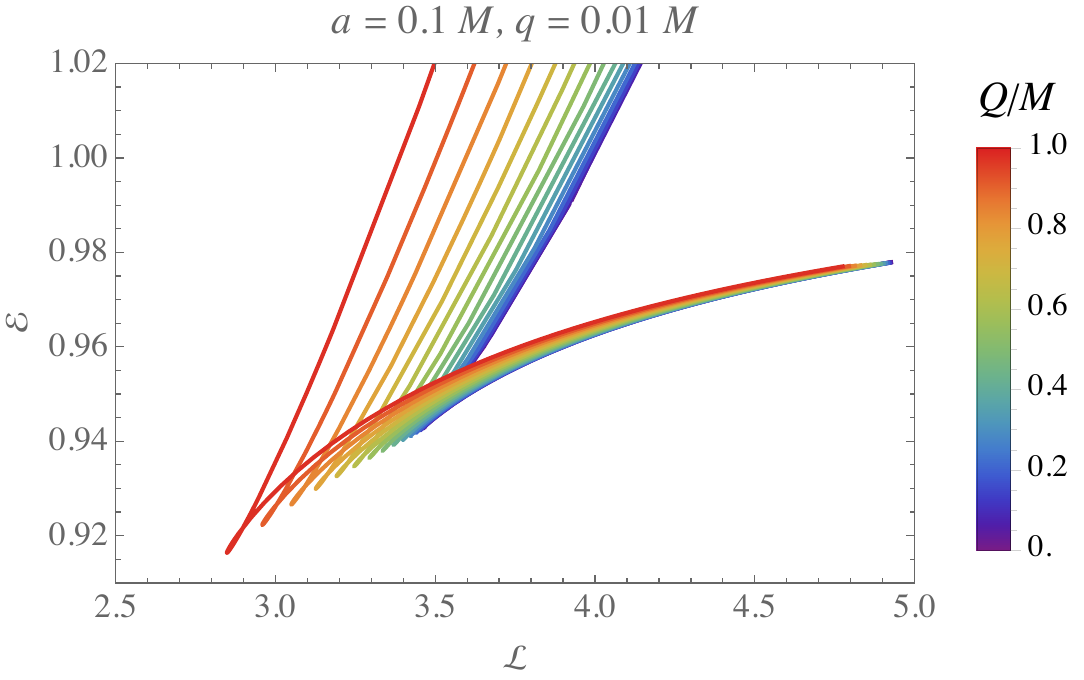} (f)
    \caption{The behavior of the conserved quantities associated with circular orbits, plotted for $a = 0.1 M$ and $\ell_p = 500 M$.  
    The first row displays the radial profiles of $\mathcal{E}(r)$ for different parameter choices, while the second row shows the corresponding behavior of $\mathcal{L}(r)$.  
    The third row illustrates the mutual dependence of $\mathcal{E}$ and $\mathcal{L}$ along circular trajectories.}
    \label{fig:EcLc}
\end{figure*}

In Fig.~\ref{fig:EcLc}, we illustrate the behavior of the conserved energy $\mathcal{E}(r)$ and angular momentum $\mathcal{L}(r)$ for circular orbits under different choices of the black hole and particle parameters.
Figures~\ref{fig:EcLc}(a)--(b) illustrate the radial dependence of the conserved specific energy $\mathcal{E}$ for charged particles moving along circular trajectories.  
When the black hole parameters are fixed and the specific charge $q$ is varied, the energy profile exhibits a clear ordering: positive values of $q$ systematically increase $\mathcal{E}$ at a given radius, while negative values of $q$ lower it.  
This behavior reflects the contribution of the electromagnetic interaction, which effectively raises (lowers) the total energy required to maintain circular motion when the particle charge is aligned (anti-aligned) with the black hole charge.

A similar trend is observed when varying the black hole charge $Q$ while keeping $q$ fixed.  
As $Q$ increases, the energy associated with circular orbits increases at all radii, indicating that stronger electromagnetic fields demand higher particle energies to counterbalance the combined gravitational and Coulomb interactions.  
In both cases, the energy profiles approach a common asymptotic behavior at large radii, where electromagnetic effects become subdominant and the spacetime geometry approaches its weak-field limit.

Figures~\ref{fig:EcLc}(c)--(d) display the corresponding radial profiles of the specific angular momentum $\mathcal{L}$.  
Increasing either the particle charge $q$ or the black hole charge $Q$ leads to a systematic enhancement of $\mathcal{L}$ for circular trajectories, particularly in the strong-field region close to the black hole.  
This indicates that charged particles require larger angular momentum to sustain circular motion as electromagnetic repulsion becomes stronger, effectively shifting circular orbits outward.

The mutual behavior of $\mathcal{E}$ and $\mathcal{L}$ is shown in Figs.~\ref{fig:EcLc}(e)--(f), where the particle charge $q$ is varied continuously in the range $-1 \le q \le 1$.  
The resulting curves reveal a smooth, monotonic correlation between energy and angular momentum, with positive values of $q$ populating regions of higher $\mathcal{E}$ and $\mathcal{L}$, while negative values of $q$ populate lower-energy and lower-angular-momentum configurations.  
This mutual dependence highlights the strong coupling between electromagnetic and gravitational effects in determining the structure of circular orbits.

{\subsection{Types of orbits}}\label{subsec:orbits_types}

\begin{figure}[t]
    \centering
    \includegraphics[width=9cm]{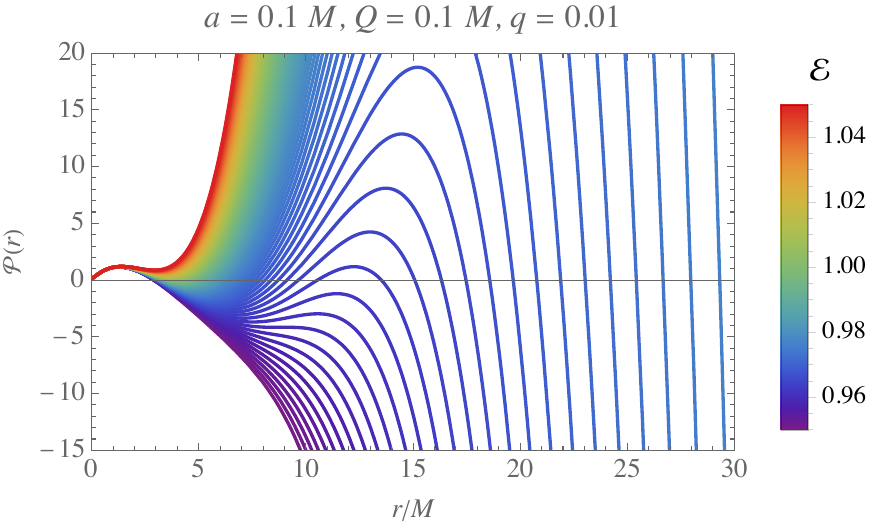}
    \caption{The radial profile of the characteristic polynomial $\mathcal{P}(r)$ for different values of $\mathcal{E}$, plotted for given initial spacetime parameter and by letting $\ell_p=500 M$ and $\mathcal{L}=4M$.}
    \label{fig:PP(r)}
\end{figure}

The equation of motion for angular trajectories reads as,
\begin{equation}
\left(\frac{dr}{d\phi}\right)^2 = \mathcal{P}(r),
    \label{eq:PP}
\end{equation}
for which, after manipulations, the characteristic polynomial can be cast as
\begin{equation}
\mathcal{P}(r)=
\rho^2\left[
\frac{\rho^2}{\mathcal{L}^2}
\left(\mathcal{E}+\frac{qQ}{\rho}\right)^2
-f(r)\left(1+\frac{\mathcal{L}^2}{\rho^2}\right)
\right],
\label{eq:PP(r)_0}
\end{equation}
where $\rho^2\equiv r^2+a^2$. 
%
In Fig. \ref{fig:PP(r)}, the radial profile of this polynomial has been plotted under different conditions.
The qualitative behavior of charged massive particle trajectories in the charged SV-AdS spacetime can be systematically classified by analyzing the real zeros of the characteristic function $\mathcal{P}(r)$.
Since the turning points of the motion are determined by the condition $\mathcal{P}(r)=0$, the number and degeneracy of its real roots provide a clear and physically transparent way to distinguish between different orbital regimes.

In fact for charged particle orbits around charged black holes, there are some general classifications of orbits, which have been listed below:

\begin{enumerate}
\item[(a)] {Three distinct real roots:}  
In this case, $\mathcal{P}(r)=0$ admits three real solutions, which we denote by $r_f$, $r_p$, and $r_a$, ordered as $r_f<r_p<r_a$.  
The smallest root $r_f$ represents a capture radius, below which the particle inevitably plunges into the black hole.  
The remaining two roots, $r_p$ and $r_a$, correspond to the periapsis and apoapsis, respectively.  
If the particle is initially located at either $r_p$ or $r_a$, or between these radii, it undergoes oscillatory radial motion, giving rise to bound or planetary-type orbits around the black hole.

\item[(b)] {Two distinct real roots:}  
When $\mathcal{P}(r)=0$ possesses two unequal real roots, denoted by $r_f$ and $r_s$, the motion is unbound.  
The radius $r_f$ again marks the capture threshold, while $r_s$ corresponds to the distance of closest approach.  
Particles approaching from infinity reach $r_s$ and are then deflected back to infinity, a process identified with the \textit{gravitational Rutherford scattering}.  
Such scattering trajectories for charged particles around charged black holes have been discussed in Refs.~\cite{villanueva_gravitational_2015,fathi_gravitational_2021}.

\item[(c)] {Two degenerate real roots:}  
If $\mathcal{P}(r)$ has a double real root, the motion corresponds to circular trajectories.  
Depending on the nature of the extremum of the effective potential, the degenerate root may represent an unstable circular orbit at $r_c$ (or critical orbit), associated with a maximum of the effective potential, or a stable circular orbit at $r_C$, corresponding to a minimum of the effective potential.

\item[(d)] {No real roots:}  
When $\mathcal{P}(r)$ has no real zeros, the particle energy exceeds the maximum of the effective potential.  
In this capture regime, there are no turning points and the particle falls directly toward the event horizon without experiencing reflection or deflection.
\end{enumerate}

It is, however, worth emphasizing that deflecting unbound trajectories with a finite turning point do not arise in the present spacetime. 
Nevertheless, a scattering-like behavior can be approached in a limiting sense. 
In the regime where $\mathcal{P}(r)$ admits three real roots, the apoapsis radius $r_a$ grows without bound as the particle energy approaches the escape threshold from below. 
In the limit $r_a \to \infty$, the motion formally resembles an unbound scattering trajectory, even though no finite turning point $r_s$ exists.

\begin{figure*}[t]
    \centering
    \includegraphics[width=5.2cm]{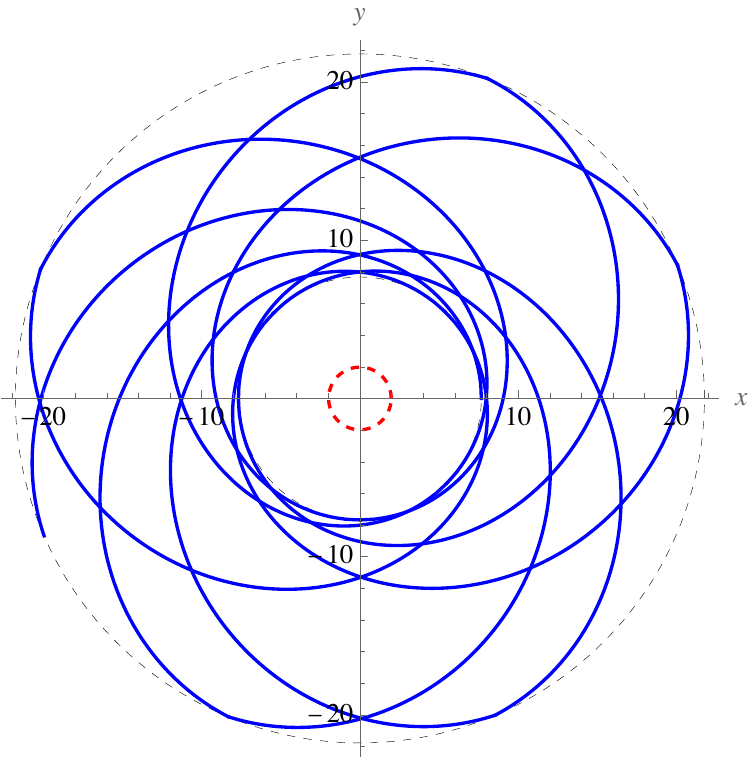} (a) \quad
    \includegraphics[width=5.2cm]{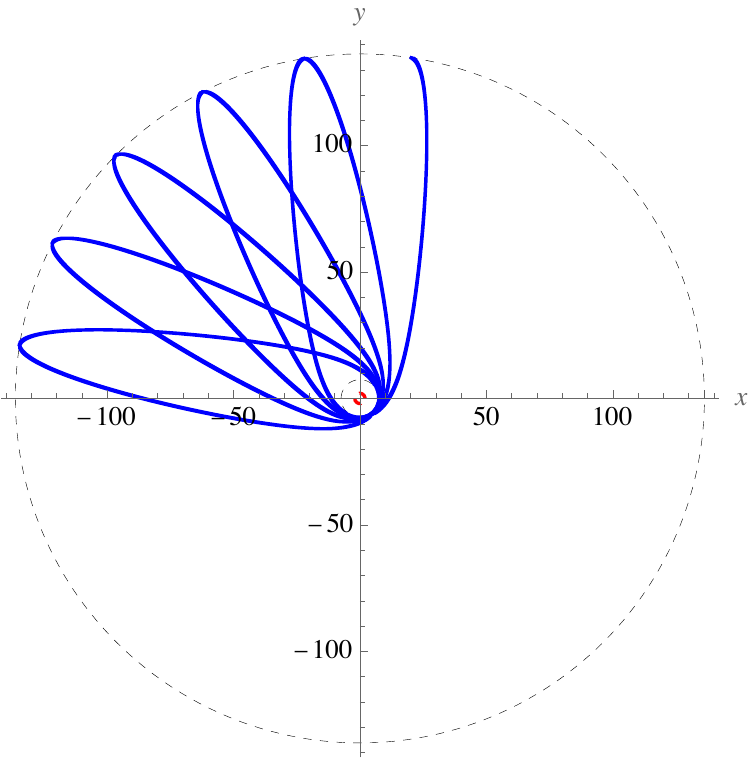} (b) \quad
    \includegraphics[width=5.2cm]{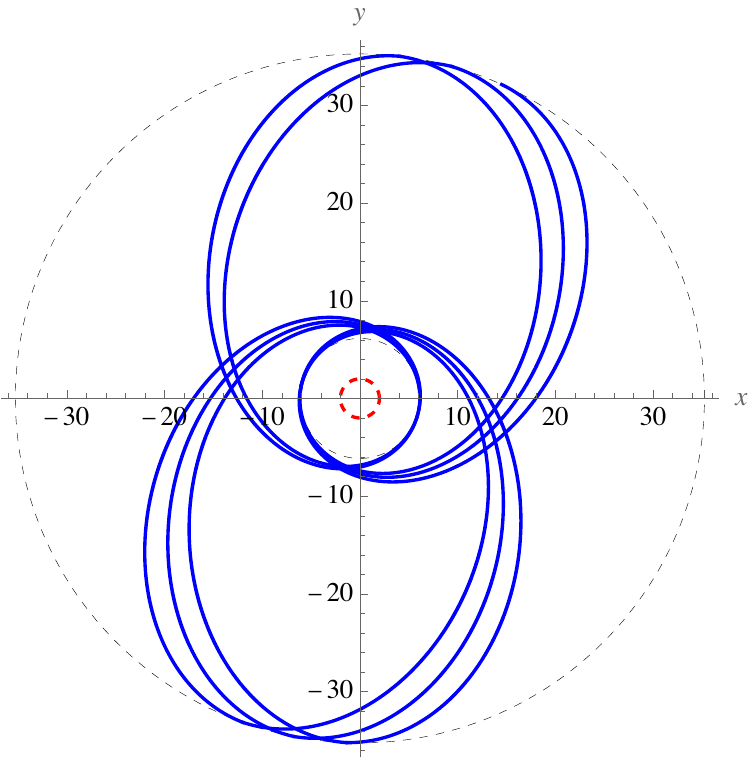} (c)
    \includegraphics[width=5.2cm]{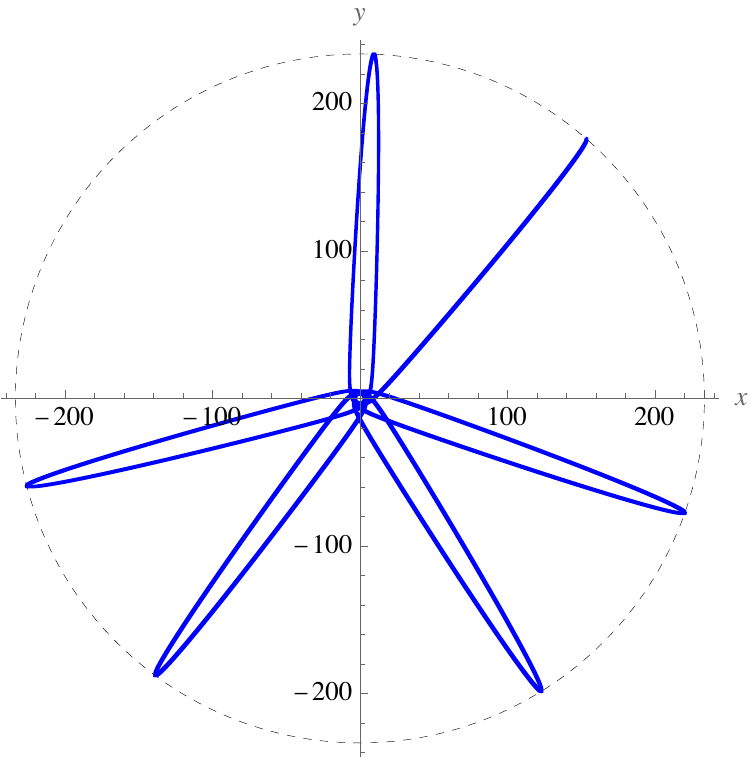} (d) \quad
    \includegraphics[width=5.2cm]{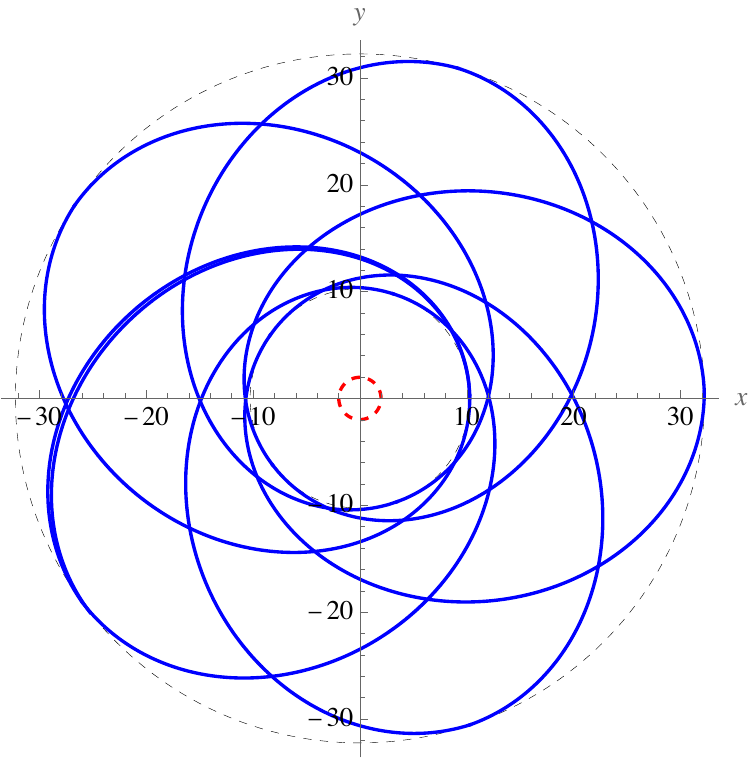} (e) \quad
    \includegraphics[width=5.2cm]{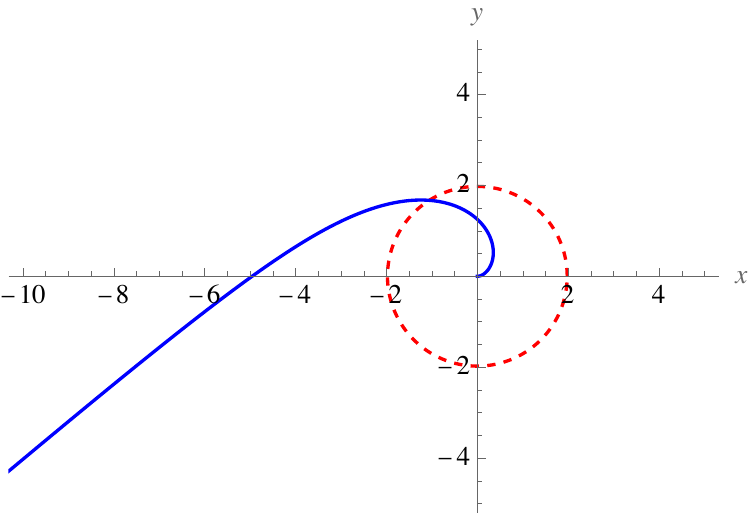} (f)
    \includegraphics[width=5.2cm]{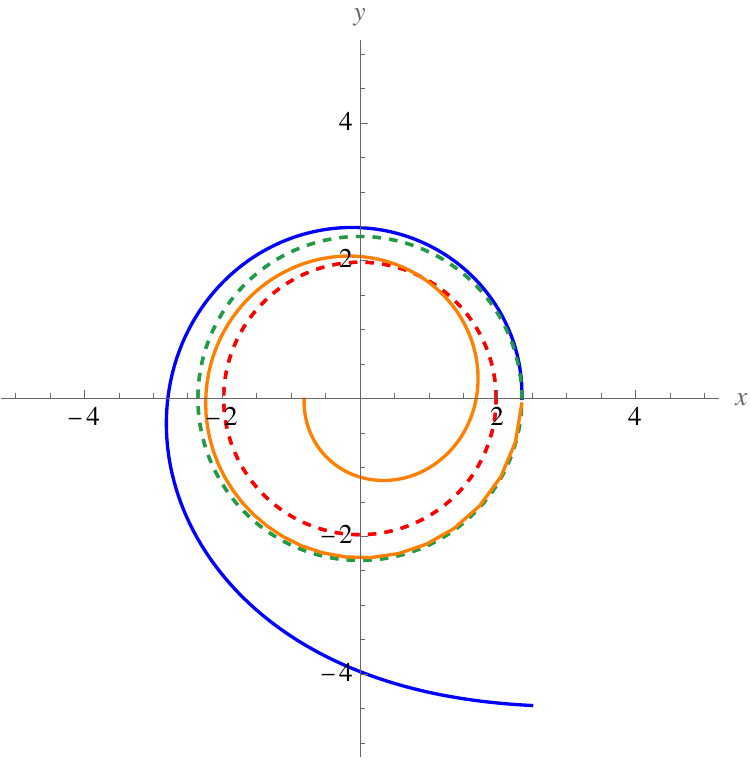} (g)
    \caption{Several representative examples of charged–particle orbits in the $(x,y)$ plane are shown, obtained by fixing $a=0.1M$ and $\ell_p=500M$. The red dashed circle at the center denotes the event–horizon radius $r_h$ for each case. Panels (a–e) correspond to planetary bound orbits, where the gray dashed circles indicate the periapsis and apoapsis radii, $r_p$ and $r_a$, respectively. Specifically, we consider: (a) $Q=0.2M$, $q=0.01M$, $\mathcal{L}=4M$, and $\mathcal{E}=0.97$; (b) $Q=0.1M$, $q=0.06M$, $\mathcal{L}=5M$, and $\mathcal{E}=1.03$; (c) $Q=0.02M$, $q=0.04M$, $\mathcal{L}=4M$, and $\mathcal{E}=0.98$; (d) $Q=1M$, $q=0.01M$, $\mathcal{L}=4M$, and $\mathcal{E}=1.1$; and (e) $Q=0.2M$, $q=0.01M$, $\mathcal{L}=4.5M$, and $\mathcal{E}=0.98$. Panel (f) illustrates the capture zone, plotted for $Q=0.2M$, $q=0.01M$, $\mathcal{L}=4M$, and $\mathcal{E}=1.5$. Panel (g) displays the critical orbits, namely the COFK, shown in blue, and the COSK, shown in orange. These are obtained for $Q=0.2M$, $q=0.01M$, and $\mathcal{L}=4M$, corresponding to the critical radius $r_c=2.351M$ (green dashed circle) and the critical energy $\mathcal{E}=1.130=\mathcal{E}_c$.}
    \label{fig:orbits}
\end{figure*}

In Fig.~\ref{fig:orbits}, several representative trajectories are presented. 
These orbits are obtained by solving the differential equation \eqref{eq:PP} to determine the pair $(r_0,\phi(r_0))$, and subsequently performing a numerical inversion of the corresponding Abelian integral to reconstruct the radial function $r(\phi)$. 
The panels included in this figure illustrate the full variety of possible equatorial orbits of charged particles around a charged SV-AdS black hole.
Note that the critical circular orbits can present two forms, which, either can escape the black hole (and in this case subject within planetary bound orbits), or fall onto the event horizon. These cases are called respectively the critical orbits of the first kind (COFK) and of the second kind (COSK).






\section{Thermodynamic Properties and Phase Structure}\label{sec:therom}

In this section, we investigate the thermodynamics of the selected black hole and analyze the combined effects of the electric charge and the regularization parameter. Black hole thermodynamics reveals a profound interplay between gravity, quantum mechanics, and statistical physics. Within this framework, black holes can be treated as thermodynamic systems characterized by well-defined quantities such as temperature, entropy, and heat capacity.

\subsubsection{Black hole Mass}

The black hole mass can be expressed in terms of the horizon radius by imposing the condition $f(r_h)=0$, where $r_h$ denotes the event horizon radius. After straightforward simplification, this condition yields
\begin{equation}
    M(r_h)=\frac{\sqrt{r^2_h+a^2}}{2}\left[1+\frac{Q^2}{r^2_h+a^2}+\frac{r^2_h+a^2}{\ell^2_p}\right].
    \label{mass}
\end{equation}

\subsubsection{Hawking temperature}

\begin{figure*}[ht!]
    \centering
    \includegraphics[width=0.46\linewidth]{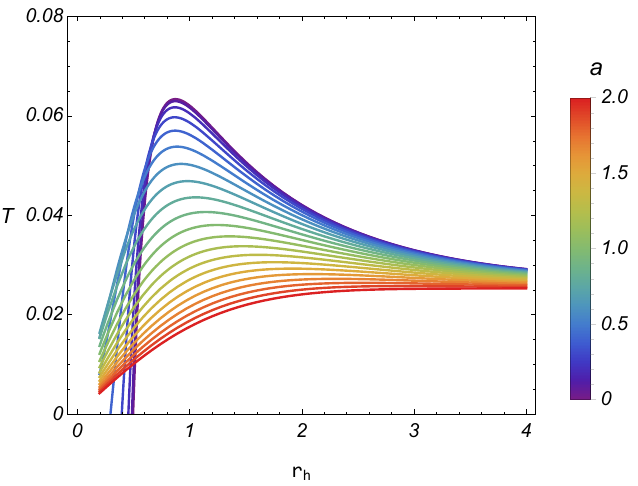}\qquad
    \includegraphics[width=0.46\linewidth]{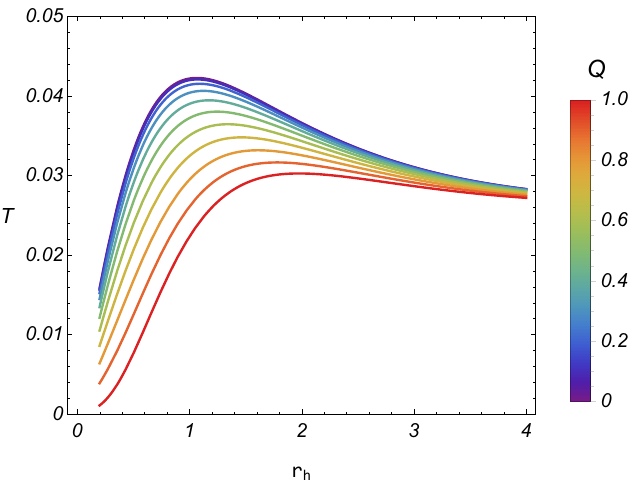}\\
    \caption{ Variation of the temperature ($T$) as a function of the horizon radius for different values of the bounce parameter $(a)$ and the electric charge $(Q)$.}
    \label{fig:temeprature}
\end{figure*}

The Hawking temperature is given by
\begin{align}
    T=\frac{f'(r_h)}{4\pi}=\frac{r_h}{4\pi}\left[\frac{1}{r_h^{2}+a^{2}} 
-\frac{Q^{2}}{(r_h^{2}+a^{2})^{2}} 
+\frac{3}{\ell_p^{2}}\right].
\label{temperature}
\end{align}
Figure \ref{fig:temeprature} displays the Hawking temperature $T$ as a function of the event horizon radius $r_h$, plotted for different values of the bounce parameter $(a)$ and the electric charge $(Q)$. In the left panel, the temperature is suppressed as $a$ increases, particularly for small black holes. A similar behavior is observed in the right panel, where increasing values of the charge parameter $Q$ lead to a reduction of the Hawking temperature.
%

\subsubsection{Entropy}

Next, we assume the validity of the first law of black hole thermodynamics and compute the entropy associated with the charged SV-AdS regular black hole spacetime. Although this assumption is nontrivial, ongoing efforts aim to establish the compatibility of the first law with regular black hole geometries \cite{murk}. For the class of black holes considered here, the first law can be written as
\begin{equation}
    dM = T dS + V dP + \Phi_{Q} dQ~.
    \label{first-law}
\end{equation}
At constant pressure, the entropy consistent with the first law can be obtained from
\begin{equation}
    S = \int \frac{dM}{T}.
\end{equation}
Evaluating the above integral yields
\begin{equation}
  \hspace{-5mm}  S=\pi\left[
r_h\sqrt{r_h^2+a^2}
+
a^2\ln\!\left(r_h+\sqrt{r_h^2+a^2}\right)
\right]~,
\label{entropy}
\end{equation}
where the horizon area is given by
\begin{align}
    &A=\int^{\pi}_{\theta=0} \int^{2\pi}_{\phi=0} \sqrt{g_{\theta\theta} g_{\phi\phi}} \, d\theta \, d\phi
    = 4 \pi (r_h^2+a^2).
    \label{area}
\end{align}
In the limiting case \(a \rightarrow 0\), corresponding to the absence of the regularization parameter, the standard Bekenstein-Hawking entropy and area relations are recovered. From Eq.~(\ref{entropy}), it is evident that the black hole entropy is independent of the charge parameter $Q$, while it explicitly depends on the bounce parameter $a$. Furthermore, the entropy $S$ given in Eq.~(\ref{entropy}) does not satisfy the standard relation $S = A/4$ for the area $A$ defined in Eq.~(\ref{area}).

\begin{figure*}[ht!]
    \centering
    \includegraphics[width=0.46\linewidth]{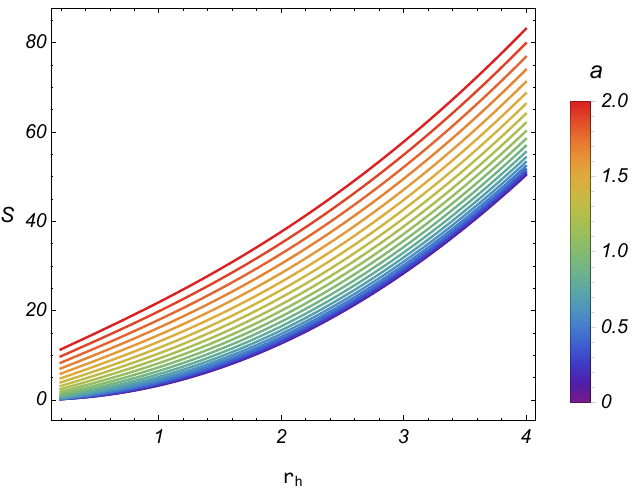}
    \caption{Variation of entropy ($S$) as a function of the horizon radius for different values of the bounce parameter $(a)$.}
    \label{fig:entropy}
\end{figure*}

Figure \ref{fig:entropy} illustrates the black hole entropy $S$ as a function of the event horizon radius $r_h$, plotted for different values of the bounce parameter. The entropy is found to increase monotonically with increasing $a$. It is worth noting that the entropy remains positive for all parameter values considered, indicating that the charged SV-AdS spacetime is physically viable, thermodynamically consistent, and compatible with both classical and quantum descriptions of black holes.
%

\subsubsection{Free energy analysis}

Free energy analysis provides essential information about the phase transition properties of a thermal system. For charged SV-AdS regular black holes, the free energy can be constructed using the thermodynamic quantities derived above. The free energy is defined as
\begin{align}
F &= M - T S \nonumber\\
  &= \frac{\sqrt{r_h^2+a^2}}{2}\left(1+\frac{Q^{2}}{r_h^2+a^2}
+\frac{r_h^2+a^2}{\ell_p^{2}}\right)\nonumber\\
&\quad -
\frac{r_h}{4}
\left(\frac{1}{r_h^2+a^2}
-\frac{Q^2}{(r_h^2+a^2)^2}
+\frac{3}{\ell_p^2}\right)\times\nonumber\\
&\quad \left[
r_h\sqrt{r_h^2+a^2}
+
a^2\ln\!\left(r_h+\sqrt{r_h^2+a^2}\right)
\right],
\label{free-energy}
\end{align}
where all quantities have their usual meanings.

In the limiting case $a \rightarrow 0$, corresponding to the absence of the regularization parameter, the expressions for the ADM mass, Hawking temperature, entropy, and free energy reduce to
\begin{align}
M(r_h)&=\frac{r_h}{2}\left(1+ \frac{Q^2}{r^2_h}+ \frac{r_h^{2}}{\ell_p^{2}}\right),\\
T(r_h)&=\frac{1}{4\pi r_h}\left(1-\frac{Q^2}{r_h^2}+\frac{3 r^2_h}{\ell_p^2}\right),\\
F(r_h)&=\frac{r_h}{4}\left(1+\frac{3Q^2}{r^2_h}-\frac{r_h^2}{\ell_p^2}\right).
\label{case}
\end{align}
These expressions coincide with the free energy of the RN-AdS black hole and further reduce to the Schwarzschild-AdS case in the limit $Q=0$ \cite{HawkingPage}.

\begin{figure*}[ht!]
    \centering
    \includegraphics[width=0.46\linewidth]{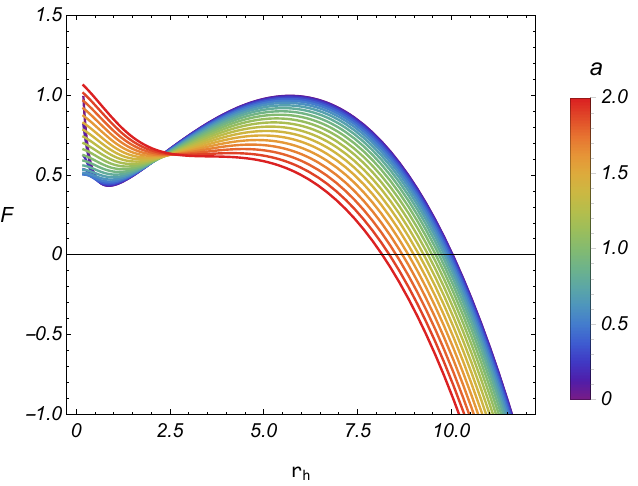}\qquad
    \includegraphics[width=0.46\linewidth]{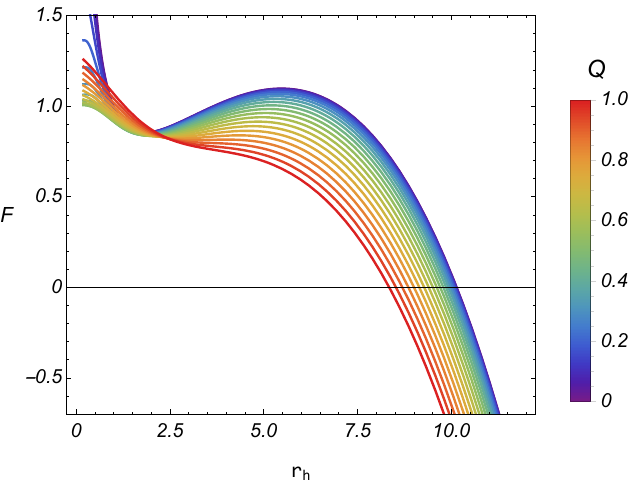}\\
    \caption{Variation of Gibbs free energy ($F$) as a function of the horizon radius for different values of the bounce parameter $(a)$ and the electric charge $(Q)$.}
    \label{fig:Gibbs}
\end{figure*}

The Helmholtz free energy $F$ as a function of the horizon radius $r_h$ for different parameter choices is shown in Fig.~\ref{fig:Gibbs}. For variations in both the bounce parameter $a$ and the charge $Q$, the Helmholtz free energy exhibits qualitatively similar behavior. In the small black hole regime, the free energy transitions from positive to negative values, signaling a change from a globally unstable to a globally stable thermodynamic phase. In contrast, large black holes remain globally stable throughout the parameter space considered.
%

\subsubsection{Specific Heat} 

\begin{figure*}[ht!]
    \centering
    \includegraphics[width=0.46\linewidth]{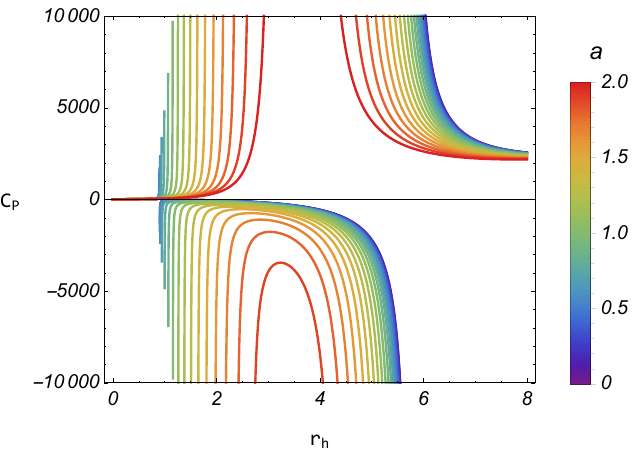}\qquad
    \includegraphics[width=0.46\linewidth]{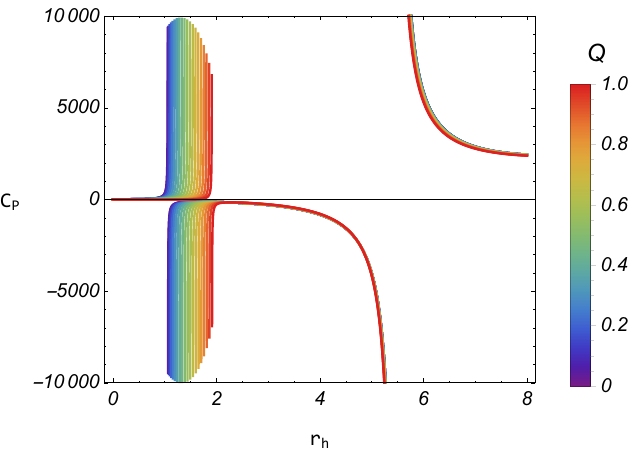}\\
    \caption{Variation of heat capacity ($C_P$) as a function of the horizon radius for different values of the bounce parameter $(a)$ and the electric charge $(Q)$.}
    \label{fig:heat}
\end{figure*}

The specific heat capacity at constant pressure is defined as
\begin{align}
    C_P &= \left(\frac{\partial M}{\partial T}\right)_P \nonumber\\
    &= 4 \pi r_h \;
\frac{
\displaystyle\frac{1}{\sqrt{r_h^2 + a^2}} - \frac{Q^2}{(r_h^2 + a^2)^{3/2}} + \frac{3 \sqrt{r_h^2 + a^2}}{\ell_p^2}
}{
\displaystyle\frac{a^2 - r_h^2}{(r_h^2 + a^2)^2} + \frac{3 Q^2 r_h^2}{(r_h^2 + a^2)^4} + \frac{3}{\ell_p^2}
}.
\label{heat}
\end{align}
In the limiting case $a=0$, the specific heat capacity reduces to
\begin{align}
    C(r_h) = 4 \pi \; \frac{\displaystyle 1 - \frac{Q^2}{r_h^2} + \frac{3 r_h^2}{\ell_p^2}}{\displaystyle -\frac{1}{r_h^2} + \frac{3 Q^2}{r_h^4} + \frac{3}{\ell_p^2}}.
    \label{heat-case}
\end{align}
Figure~\ref{fig:heat} illustrates the behavior of the heat capacity as a function of the horizon radius $r_h$ for different regions of the parameter space. The heat capacity $C_P$ exhibits both positive and negative values, indicating the coexistence of thermodynamically stable and unstable black hole phases. The divergences of $C_P$ signal second-order phase transitions separating these regimes. Increasing the bounce parameter $a$ shifts the critical points toward larger horizon radii and alters the extent of the thermodynamically stable phase. Similarly, variations in the electric charge $Q$ significantly influence the location of the divergence points. Hence, the presence of extended regions with $C_P>0$ confirms the existence of thermodynamically stable black hole configurations over a wide range of parameters.
%

\section{Modified first law of thermodynamics} \label{sec:therom2}

In the extended phase space approach to black hole thermodynamics, the cosmological constant $\Lambda = -3/\ell_p^2$ is treated as a thermodynamic variable and is identified with the pressure defined by $P = -\Lambda/(8\pi)$. For AdS spacetimes, where $\Lambda < 0$, this definition corresponds to a positive thermodynamic pressure. Within this framework, the ADM mass $M$ of the black hole is interpreted as the enthalpy of the system rather than its internal energy. Consequently, the first law of charged black hole thermodynamics is generalized to include the $V dP + \Phi_Q dQ$ contributions \cite{Kubiznak2012,Dolan2011,Kastor2009,Gunasekaran2012}, where
\begin{equation}
    V = \left(\frac{\partial M}{\partial P}\right)_{S,Q}.
    \label{volume}
\end{equation}
Here, $\Phi_Q$ denotes the electric potential.

In terms of the pressure $P$, the ADM mass $M$ can be rewritten as
\begin{equation}
    M(r_h)=\frac{\sqrt{r_h^2+a^2}}{2}\left[1+\frac{Q^2}{r_h^2+a^2}+\frac{8\pi}{3}(r_h^2+a^2) P\right].
    \label{mass-correction}
\end{equation}
Accordingly, the thermodynamic volume is obtained as
\begin{equation}
    V=\frac{4\pi}{3} (r_h^2+a^2)^{3/2}.
    \label{volume-correction}
\end{equation}
Similarly, the electric potential associated with the charge $Q$ is given by
\begin{equation}
    \Phi_{Q}=\left(\frac{\partial M}{\partial Q}\right)_{S,P}=\frac{Q}{\sqrt{r_h^2+a^2}}.
    \label{potential}
\end{equation}
One can readily verify that the thermodynamic quantities satisfy the Smarr relation
$2(TS - PV) + Q \Phi_Q = 2 M(r_h)$,
which confirms the consistency of the extended phase space formulation for this charged black hole solution.

\section{Equation of state}\label{sec:therom3}

In extended phase space thermodynamics, the cosmological constant $\Lambda$ is promoted to a thermodynamic variable and is naturally interpreted as the pressure $P$ of the black hole system, defined by
\begin{equation}
\frac{1}{\ell_p^2} = -\frac{\Lambda}{3} = 8\pi P .
\end{equation}
Within this framework, the black hole mass is reinterpreted as the enthalpy rather than the internal energy of the system. Treating $\Lambda$ as a thermodynamic variable leads to a richer thermodynamic structure, including Van der Waals-like phase transitions, critical phenomena, and a consistent formulation of the first law and the Smarr relation \cite{Kastor2009,Dolan2011,Kubiznak2012,Cvetic2011}.

The Hawking temperature expressed in terms of the pressure $P$ is given by
\begin{equation}
T = \frac{r_h}{4 \pi} \left( \frac{1}{r_h^2 + a^2} - \frac{Q^2}{(r_h^2 + a^2)^2} + 8 \pi P \right).
\label{ss1}
\end{equation}
Rewriting the above expression in terms of $P$ yields the equation of state
\begin{align}
P = \frac{T}{2 r_h} - \frac{1}{8 \pi (r_h^2 + a^2)} + \frac{Q^2}{8 \pi (r_h^2 + a^2)^2}.
\end{align}
Substituting $r_h = \sqrt{\left( {3 V}/{4 \pi} \right)^{2/3} - a^2}$ into the equation of state, one obtains
\begin{equation}
P = \frac{T}{2 r_h} - \frac{1}{\displaystyle 8 \pi \left( \frac{3 V}{4 \pi} \right)^{2/3}} + \frac{Q^2}{\displaystyle 8 \pi \left( \frac{3 V}{4 \pi} \right)^{4/3}}.
\end{equation}
The main thermodynamic features and critical behavior implied by the equation of state can be summarized as follows:
\begin{itemize}
    \item The presence of a nonvanishing bounce parameter $a \neq 0$ introduces a geometric correction to the effective horizon size, which for fixed $T$ and $V$ leads to a reduction of the pressure compared to the standard RN-AdS case.
    \item The equation of state admits critical points determined by the inflection point conditions
    \begin{align}
    &\left(\frac{\partial P}{\partial V}\right)_T = 0, \quad
    &\left(\frac{\partial^2 P}{\partial V^2}\right)_T = 0.
    \end{align}
    \item In the limiting case $a \to 0$, the standard RN-AdS equation of state is recovered,
    \begin{align}
    P = \frac{T}{2 r_h} - \frac{1}{8 \pi r_h^2} + \frac{Q^2}{8 \pi r_h^4}.
    \end{align}
    \item The resulting equation of state exhibits a clear analogy with the Van der Waals fluid, with the parameter $a$ modifying the critical behavior and phase structure while preserving the overall thermodynamic universality.
\end{itemize}

\section{Conclusions} \label{conclu}

Singularities in general relativity are unavoidable when the standard energy conditions for ordinary matter are satisfied. Their presence, however, signals the breakdown of general relativity as a complete classical theory of gravity and strongly motivates the exploration of alternative frameworks capable of regularizing such divergences. The SV-regularization scheme provides a systematic and bottom-up approach to resolving the central singularity of black holes. By implementing this procedure, one obtains a spacetime geometry that is free of singularities, with regularization effects that manifest themselves at both the classical and thermodynamic levels.

In this work, we applied the SV-regularization procedure to a charged black hole in AdS spacetime and carried out a detailed investigation of its classical and thermodynamic properties. From a classical perspective, the resulting spacetime is regular everywhere and free from central singularities for all nonzero values of the SV-regularization parameter $a$. We analyzed null geodesics in this background, focusing on the effective potential and photon trajectories, and examined the associated dynamical behavior. Within the weak-field approximation, we computed the deflection angle of light and demonstrated how the regularization parameter $a$ modifies photon bending. We also studied the black hole shadow and compared our theoretical predictions with the EHT observations of M87* and Sgr A*, allowing us to place meaningful constraints on the total charge $Q$ of the black hole. In contrast, constraining the SV-regularization parameter $a$ remains challenging, as the outer shadow radius alone does not provide sufficiently strong sensitivity. Tighter bounds on this parameter likely require more stringent observational probes, such as the dynamics of the S2 star at the Galactic center and its associated quasi-periodic oscillations. In addition, we investigated the motion of charged particles in the vicinity of the black hole, with particular emphasis on circular orbits, and showed how the regularization parameter modifies the corresponding orbital structure.

On the thermodynamic side, assuming that the regularized geometry satisfies the first law of black hole thermodynamics, we derived the entropy of the charged SV-AdS black hole. The resulting entropy exhibits a nontrivial dependence on the regularization parameter $a$, while the standard Bekenstein-Hawking entropy is recovered smoothly in the limit $a \to 0$. The presence of $a$ also affects other thermodynamic quantities, most notably inducing deviations in the Hawking temperature relative to standard charged AdS black holes. The thermodynamic phase structure is significantly altered by the SV-regularization. In the conventional AdS scenario, a Hawking-Page phase transition occurs between thermal AdS space and a black hole phase. When the regularization parameter is nonzero, this structure is qualitatively modified: the thermal AdS phase disappears, and a black hole phase exists across the entire temperature range. An analysis of the free energy reveals a small-to-large black hole phase transition as the Hawking temperature varies. The critical behavior of this transition closely resembles that of a Van der Waals fluid, featuring a well-defined critical point that depends explicitly on the regularization parameter $a$.

In summary, the SV-regularization framework introduces profound modifications to both the classical and thermodynamic properties of charged AdS black holes. By eliminating the central singularity, modifying photon trajectories, particle dynamics, and black hole shadows, and fundamentally reshaping the thermodynamic phase structure, this approach provides a consistent and physically rich setting for the study of regular charged black holes in AdS spacetimes.

\section*{Acknowledgment}

F.A. acknowledges the Inter University Center for Astronomy and Astrophysics (IUCAA), Pune, India for granting visiting associateship. The work of M.F. has been supported by Universidad Central de Chile through the project No. PDUCEN20240008.

\section*{Data Availability Statement}

No data is associated with this manuscript [Author's comment: No new data were generated or created in this study].

\section*{Code/Software}

No new code/software were developed in this manuscript [Author's comment: No code/software were developed or created in this study].

\bibliographystyle{ieeetr}
\bibliography{1biblio}

\begin{thebibliography}{10}

\bibitem{EHTL1}
K.~Akiyama and et~al. (EHT~Collaboration), ``First m87 event horizon telescope results. i. the shadow of the supermassive black hole,'' {\em Astrophys. J. Lett.}, vol.~875, p.~L1, 2019.

\bibitem{EHTL4}
K.~Akiyama and et~al. (EHT~Collaboration), ``First m87 event horizon telescope results. iv. imaging the central supermassive black hole,'' {\em Astrophys. J. Lett.}, vol.~875, p.~L4, 2019.

\bibitem{EHTL6}
K.~Akiyama and et~al. (EHT~Collaboration), ``First m87 event horizon telescope results. vi. the shadow and mass of the central black hole,'' {\em Astrophys. J. Lett.}, vol.~875, p.~L6, 2019.

\bibitem{EHTL12}
K.~Akiyama and et~al. (EHT~Collaboration), ``First sagittarius a* event horizon telescope results. xii. the central supermassive black hole shadow,'' {\em Astrophys. J. Lett.}, vol.~930, p.~L12, 2022.

\bibitem{EHTL25}
K.~Akiyama and et~al. (EHT~Collaboration), ``Sagittarius a* event horizon telescope results. xxv. observational properties of the galactic center black hole,'' {\em Astrophys. J. Lett.}, vol.~964, p.~L25, 2024.

\bibitem{EHTL26}
K.~Akiyama and et~al. (EHT~Collaboration), ``Sagittarius a* event horizon telescope results. xxvi. constraints on accretion and spin of the central black hole,'' {\em Astrophys. J. Lett.}, vol.~964, p.~L26, 2024.

\bibitem{Bardeen1973}
J.~M. Bardeen, B.~Carter, and S.~W. Hawking, ``The four laws of black hole mechanics,'' {\em Commun. Math. Phys.}, vol.~31, p.~161, 1973.

\bibitem{Wald2001}
R.~M. Wald, ``The thermodynamics of black holes,'' {\em Living Rev. Relativity}, vol.~4, p.~6, 2001.

\bibitem{Hawking1975}
S.~W. Hawking, ``Particle creation by black holes,'' {\em Commun. Math. Phys.}, vol.~43, p.~199, 1975.

\bibitem{Hawking1976}
S.~W. Hawking, ``Black holes and thermodynamics,'' {\em Commun. Math. Phys.}, vol.~46, p.~206, 1976.

\bibitem{Bekenstein1972}
J.~D. Bekenstein, ``Black holes and the second law,'' {\em Lett. Nuovo Cimento}, vol.~4, p.~737, 1972.

\bibitem{Bekenstein1973}
J.~D. Bekenstein, ``Black holes and entropy,'' {\em Phys. Rev. D}, vol.~7, p.~2333, 1973.

\bibitem{Penrose}
R.~Penrose, ``Gravitational {Collapse} and {Space}-{Time} {Singularities},'' {\em Physical Review Letters}, vol.~14, pp.~57--59, Jan. 1965.

\bibitem{Penrose1}
R.~Penrose, ``“{Golden} {Oldie}”: {Gravitational} {Collapse}: {The} {Role} of {General} {Relativity},'' {\em General Relativity and Gravitation}, vol.~34, pp.~1141--1165, July 2002.

\bibitem{kiefer01}
C.~Kiefer, {\em Quantum gravity}.
\newblock No.~176 in International series of monographs on physics, Oxford: Oxford University Press, fourth edition~ed., 2025.

\bibitem{kiefer02}
C.~{Kiefer}, ``{Quantum gravity -- an unfinished revolution},'' {\em arXiv e-prints}, p.~arXiv:2302.13047, Feb. 2023.

\bibitem{Bardeen}
J.~{Bardeen}, ``{Non-singular general relativistic gravitational collapse},'' in {\em Proceedings of the 5th International Conference on Gravitation and the Theory of Relativity}, p.~87, Sept. 1968.

\bibitem{reg2}
E.~Ayón-Beato and A.~García, ``Regular {Black} {Hole} in {General} {Relativity} {Coupled} to {Nonlinear} {Electrodynamics},'' {\em Physical Review Letters}, vol.~80, pp.~5056--5059, June 1998.

\bibitem{reg3}
A.~Borde, ``Open and closed universes, initial singularities, and inflation,'' {\em Physical Review D}, vol.~50, pp.~3692--3702, Sept. 1994.

\bibitem{reg4}
S.~A. Hayward, ``Formation and {Evaporation} of {Nonsingular} {Black} {Holes},'' {\em Physical Review Letters}, vol.~96, p.~031103, Jan. 2006.

\bibitem{reg5}
V.~P. Frolov, ``Notes on nonsingular models of black holes,'' {\em Physical Review D}, vol.~94, p.~104056, Nov. 2016.

\bibitem{reg6}
C.~Lan, H.~Yang, Y.~Guo, and Y.-G. Miao, ``Regular {Black} {Holes}: {A} {Short} {Topic} {Review},'' {\em International Journal of Theoretical Physics}, vol.~62, p.~202, Sept. 2023.

\bibitem{SV}
A.~Simpson and M.~Visser, ``Black-bounce to traversable wormhole,'' {\em Journal of Cosmology and Astroparticle Physics}, vol.~2019, pp.~042--042, Feb. 2019.

\bibitem{Kumar2025}
N.~Kumar, A.~Srivastav, and P.~Channuie, ``Simpson-visser-ads black holes: Thermodynamics and binary merger,'' {\em arXiv preprint}, 2025.

\bibitem{Wu2000}
X.~N. Wu, ``Multicritical phenomena of reissner-nordström anti–de sitter black holes,'' {\em Phys. Rev. D}, vol.~62, p.~124023, 2000.

\bibitem{Synge1966}
J.~L. Synge, ``The {Escape} of {Photons} from {Gravitationally} {Intense} {Stars},'' {\em Monthly Notices of the Royal Astronomical Society}, vol.~131, pp.~463--466, Feb. 1966.

\bibitem{Luminet1979}
J.-P. {Luminet}, ``{Image of a spherical black hole with thin accretion disk.},'' {\em Astron. Astrophys.}, vol.~75, pp.~228--235, May 1979.

\bibitem{VirbhadraEllis2000}
K.~S. Virbhadra and G.~F.~R. Ellis, ``Schwarzschild black hole lensing,'' {\em Physical Review D}, vol.~62, p.~084003, Sept. 2000.

\bibitem{EHT2019I}
{The Event Horizon Telescope Collaboration}, Akiyama, {\em et~al.}, ``First {M87} {Event} {Horizon} {Telescope} {Results}. {I}. {The} {Shadow} of the {Supermassive} {Black} {Hole},'' {\em The Astrophysical Journal Letters}, vol.~875, p.~L1, Apr. 2019.

\bibitem{event_horizon_telescope_collaboration_first_2022}
{Event Horizon Telescope Collaboration}, K.~Akiyama, {\em et~al.}, ``First {Sagittarius} {A}* {Event} {Horizon} {Telescope} {Results}. {I}. {The} {Shadow} of the {Supermassive} {Black} {Hole} in the {Center} of the {Milky} {Way},'' {\em The Astrophysical Journal Letters}, vol.~930, p.~L12, May 2022.

\bibitem{ALBADAWI2025185}
A.~Al-Badawi and F.~Ahmed, ``A new black hole coupled with nonlinear electrodynamics surrounded by quintessence: Thermodynamics, geodesics, and regge–wheeler potential,'' {\em Chinese Journal of Physics}, vol.~94, pp.~185--203, 2025.

\bibitem{AHMED2025101925}
F.~Ahmed, A.~Al-Badawi, and İzzet Sakallı, ``Probing quantum gravity effects: Geodesic structure and thermodynamics of deformed schwarzschild ads black holes surrounded by cosmic strings,'' {\em Physics of the Dark Universe}, vol.~48, p.~101925, 2025.

\bibitem{AHMED2025116951}
F.~Ahmed, A.~Al-Badawi, and İzzet Sakallı, ``Exploring geodesics, quantum fields and thermodynamics of schwarzschild-ads black hole with a global monopole in non-commutative geometry,'' {\em Nuclear Physics B}, vol.~1017, p.~116951, 2025.

\bibitem{AHMED2025101988}
F.~Ahmed, A.~Al-Badawi, and İzzet Sakallı, ``Photon spheres, gravitational lensing/mirroring, and greybody radiation in deformed ads-schwarzschild black holes with phantom global monopole,'' {\em Physics of the Dark Universe}, vol.~49, p.~101988, 2025.

\bibitem{bozza_gravitational_2010}
V.~Bozza, ``Gravitational lensing by black holes,'' {\em General Relativity and Gravitation}, vol.~42, pp.~2269--2300, Sept. 2010.

\bibitem{Synge:1966}
J.~L. Synge, ``The escape of photons from gravitationally intense stars,'' {\em Monthly Notices of the Royal Astronomical Society}, vol.~131, pp.~463--466, 1966.

\bibitem{Cunningham:1972}
C.~T. Cunningham and J.~M. Bardeen, ``The optical appearance of a star orbiting an extreme kerr black hole,'' {\em Astrophysical Journal Letters}, vol.~173, pp.~L137--L142, 1972.

\bibitem{Bardeen:1973a}
J.~M. {Bardeen}, W.~H. {Press}, and S.~A. {Teukolsky}, ``{Rotating Black Holes: Locally Nonrotating Frames, Energy Extraction, and Scalar Synchrotron Radiation},'' {\em Astrophys. J}, vol.~178, pp.~347--370, Dec. 1972.

\bibitem{Luminet:1979nyg}
J.~P. Luminet, ``{Image of a spherical black hole with thin accretion disk},'' {\em Astron. Astrophys.}, vol.~75, pp.~228--235, 1979.

\bibitem{zeldovich_relativistic_1966}
Y.~B. Zeldovich and I.~D. Novikov, {\em Relativistic Astrophysics, Volume 1: Stars and Relativity}.
\newblock University of Chicago Press, 1971.
\newblock Original Russian Edition: 1966.

\bibitem{Bardeen:1973b}
J.~M. Bardeen, ``Black holes: Les astres occlus,'' in {\em Proceedings of the Les Houches Summer School} (C.~DeWitt and B.~S. DeWitt, eds.), pp.~215--239, Gordon and Breach, 1973.

\bibitem{Chandrasekhar:1998}
S.~Chandrasekhar, {\em The mathematical theory of black holes}.
\newblock Oxford classic texts in the physical sciences, Oxford University Press, 1998.

\bibitem{luminet_seeing_2018}
J.-P. Luminet, ``Seeing black holes: from astronomical curiosity to observational reality,'' {\em The European Physical Journal H}, vol.~43, no.~2, pp.~293--329, 2018.

\bibitem{Falcke_2000}
H.~Falcke, F.~Melia, and E.~Agol, ``Viewing the shadow of the black hole at the galactic center,'' {\em Astrophysical Journal Letters}, vol.~528, pp.~L13--L16, 2000.

\bibitem{johannsen2010}
T.~Johannsen and D.~Psaltis, ``{Testing the No-Hair Theorem with Observations in the Electromagnetic Spectrum: II. Black-Hole Images},'' {\em Astrophys. J.}, vol.~718, pp.~446--454, 2010.

\bibitem{johnson_universal_2020}
M.~D. Johnson {\em et~al.}, ``{Universal interferometric signatures of a black hole{\textquoteright}s photon ring},'' {\em Sci. Adv.}, vol.~6, no.~12, p.~eaaz1310, 2020.

\bibitem{bambi_testing_2019}
C.~Bambi, K.~Freese, S.~Vagnozzi, and L.~Visinelli, ``Testing the rotational nature of the supermassive object {M87}* from the circularity and size of its first image,'' {\em Physical Review D}, vol.~100, p.~044057, Aug. 2019.

\bibitem{villanueva_gravitational_2015}
J.~R. Villanueva and M.~Olivares, ``Gravitational {Rutherford} scattering and {Keplerian} orbits for electrically charged bodies in heterotic string theory,'' {\em The European Physical Journal C}, vol.~75, p.~562, Nov. 2015.

\bibitem{fathi_gravitational_2021}
M.~Fathi, M.~Olivares, and J.~R. Villanueva, ``Gravitational {Rutherford} scattering of electrically charged particles from a charged {Weyl} black hole,'' {\em The European Physical Journal Plus}, vol.~136, p.~420, Apr. 2021.

\bibitem{murk}
S.~Murk and I.~Soranidis, ``Regular black holes and the first law of black hole mechanics,'' {\em Physical Review D}, vol.~108, p.~044002, Aug. 2023.

\bibitem{HawkingPage}
S.~W. Hawking and D.~N. Page, ``Thermodynamics of black holes in anti-de {Sitter} space,'' {\em Communications in Mathematical Physics}, vol.~87, pp.~577--588, Dec. 1983.

\bibitem{Kubiznak2012}
D.~Kubiz{\v n}{\'a}k and R.~B. Mann, ``P--v criticality of charged ads black holes,'' {\em JHEP}, vol.~1207, p.~033, 2012.

\bibitem{Dolan2011}
B.~P. Dolan, ``The cosmological constant and the black hole equation of state,'' {\em Class. Quant. Grav.}, vol.~28, p.~125020, 2011.

\bibitem{Kastor2009}
D.~Kastor, S.~Ray, and J.~Traschen, ``Enthalpy and the mechanics of ads black holes,'' {\em Class. Quant. Grav.}, vol.~26, p.~195011, 2009.

\bibitem{Gunasekaran2012}
S.~Gunasekaran, D.~Kubiz{\v n}{\'a}k, and R.~B. Mann, ``Extended phase space thermodynamics for charged and rotating black holes and born-infeld vacuum polarization,'' {\em JHEP}, vol.~1211, p.~110, 2012.

\bibitem{Cvetic2011}
M.~Cveti\v{c}, G.~W. Gibbons, D.~Kubiz\v{n}\'ak, and C.~N. Pope, ``Black hole enthalpy and an entropy inequality for the thermodynamic volume,'' {\em Phys. Rev. D}, vol.~84, p.~024037, 2011.

\end{thebibliography}

\end{document}